\documentclass[12pt,preprint]{aastex}

\newcommand{\mus}{\mbox{$\mu$s}}
\newcommand{\muas}{\mbox{$\mu$as}}
\newcommand{\muasyr}{\mbox{$\mu$as~yr$^{-1}$}}
\newcommand{\muasyryr}{\mbox{$\mu$as~yr$^{-2}$}}
\newcommand{\masyr}{\mbox{mas~yr$^{-1}$}}
\newcommand{\uv}{\mbox{$u$-$v$}}
\newcommand{\GPB}{\mbox{\em GP-B}}

\newcommand{\kmsMpc}{\mbox{km s$^{-1}$ Mpc$^{-1}$}}
\newcommand{\Jb}{\mbox{Jy bm$^{-1}$}}
\newcommand{\mJb}{\mbox{mJy bm$^{-1}$}}
\newcommand{\Ra}[4]{\mbox{${#1}^{\rm h} \; {#2}^{\rm m} \; {#3}\fs{#4} $}}
\newcommand{\dec}[4]{\mbox{${#1}\arcdeg \; {#2}\arcmin \; {#3}\farcs{#4} $}}\newcommand{\al}{\mbox{$\alpha$}}
\newcommand{\de}{\mbox{$\delta$}}
\newcommand{\mua}{\mbox{$\mu_{\alpha}$}}
\newcommand{\mud}{\mbox{$\mu_{\delta}$}}
\newcommand{\dotmua}{\mbox{$\dot\mu_{\alpha}$}}
\newcommand{\dotmud}{\mbox{$\dot\mu_{\delta}$}}
\newcommand{\chinu}{\mbox{$\chi^2_\nu / \nu$}}

\slugcomment{Accepted to the Astrophysical Journal Supplement Series}

\shortauthors{Bartel  et al.}
\shorttitle{GP-B Astrometry (III)}

\begin{document}

\title{VLBI for {\em Gravity Probe B}. III. A Limit on the Proper Motion of the
``Core'' of the Quasar 3C 454.3}

\author{N. Bartel\altaffilmark{1}, M. F. Bietenholz\altaffilmark{1,2},
D. E. Lebach\altaffilmark{3}, J. I. Lederman\altaffilmark{4},
L. Petrov\altaffilmark{5},
R. R. Ransom\altaffilmark{1,6}, M. I. Ratner\altaffilmark{3}, and
I. I. Shapiro\altaffilmark{3}}

\altaffiltext{1}{Department of Physics and Astronomy, York University,
4700 Keele Street, Toronto, ON M3J 1P3, Canada}

\altaffiltext{2}{Now also at Hartebeesthoek Radio Astronomy Observatory,
PO Box 443, Krugersdorp 1740, South Africa}

\altaffiltext{3}{Harvard-Smithsonian Center for Astrophysics, 60
Garden Street, Cambridge, MA 02138, USA}

\altaffiltext{4}{York University, Centre for Research in Earth and
Space Sciences, 4700 Keele Street, Toronto, ON M3J 1P3, Canada}

\altaffiltext{5}{Astrogeo Center, 7312 Sportsman Drive, Falls Church, VA 22043, USA}

\altaffiltext{6}{Now at Okanagan College, 583 Duncan Avenue
West, Penticton, B.C., V2A 2K8, Canada and also at the National
Research Council of Canada, Herzberg Institute of Astrophysics,
Dominion Radio Astrophysical Observatory, P.O. Box 248, Penticton,
B.C., V2A 6K3, Canada}


\begin{abstract}

We made VLBI observations at 8.4 GHz between 1997 and 2005 to estimate
the coordinates of the ``core'' component of the superluminal quasar,
3C~454.3, the ultimate reference point in the distant universe for the
NASA/Stanford Gyroscope Relativity Mission, {\em Gravity Probe B}. These
coordinates are determined relative to those of the brightness peaks of
two other compact extragalactic sources, B2250+194 and B2252+172,
nearby on the sky, and within a celestial reference frame (CRF),
defined by a large suite of compact extragalactic radio sources, and
nearly identical to the International Celestial Reference Frame 2
(ICRF2). We find that B2250+194 and B2252+172 are stationary relative
to each other, and also in the CRF, to within 1$\sigma$ upper limits
of 15 and 30~\muasyr\ in $\alpha$ and $\delta$, respectively.
The core of 3C~454.3 appears to jitter in its position along the jet direction over 
$\sim0.2$~mas, likely due to activity close to the putative supermassive black hole nearby, but
on average is stationary in the CRF within 1$\sigma$ upper limits on its proper
motion of 39~\muasyr\ ($1.0 c$) and 30~\muasyr\ ($0.8 c$) in \al\ and
\de, respectively, for the period 2002 $-$ 2005. Our corresponding limit over 
the longer interval, 1998 $-$ 2005, of more importance to \GPB, is 46 and 
56~\muasyr\ in \al\ and \de, respectively. Some of 3C~454.3's jet components show
significantly superluminal motion with speeds of up to
$\sim$200~\muasyr\ or $5 c$ in the CRF.
The
core of 3C~454.3 thus provides for {\em Gravity Probe B} a
sufficiently stable reference in the distant universe.

\end{abstract}

\section{Introduction}
\label{sintro}

{\em Gravity Probe B} (\GPB) is the spaceborne relativity experiment
developed by NASA and Stanford University to test two predictions of
general relativity. The experiment used four superconducting
gyroscopes, contained in a low-altitude, polar orbiting spacecraft, to
measure the geodetic effect and the much smaller frame-dragging
effect. According to general relativity, each of these effects induces
precessions of the gyroscopes in planes perpendicular to each
other. For the geodetic effect, which depends directly on the Earth's mass, 
the predicted precession is 6.6 arcsec yr$^{-1}$ and for the frame-dragging 
effect, which depends directly on the angular momentum of the Earth, it is 39~\masyr. \GPB\ was
expected to measure each precession with a standard error $\leq$0.5
\masyr\ relative to the distant universe.  Because of technical
limitations, the spacecraft could not measure the precessions directly
relative to the distant universe but only to an optically bright star, the guide
star, chosen to be IM Pegasi (\objectname[]{HR 8703}).  We must
therefore determine IM~Peg's proper motion relative to the distant universe,
which is, for our purposes, best represented by
extragalactic radio sources.

For our part of the \GPB\ project, we determined the coordinates
and the proper motion of the guide star in the radio relative to the ``core'' of
the quasar 3C~454.3 (B2251+158). This core was tied to two other radio
sources, which are compact, extragalactic, nearby to it on the sky, and also tied to a celestial 
reference frame (CRF) defined by a large suite of extragalactic sources.
These ties are the main subject of this paper. Most important for \GPB, 
of course, is the bound that we place on the proper motion of the core, 
which serves as the principal reference for determining the proper motion of 
 \objectname[]{IM Peg}. The \GPB\ project needs the proper motion of the optical 
source in \objectname[]{IM Peg}. The radio source in \objectname[]{IM Peg}, 
however, moves erratically with respect to the optical source. In order to be 
able to average as well as feasible over the erratic motion, we place our VLBI 
limit on the motion of the core of 3C~454.3 over as long a period as feasible. 
By contrast, for astrophysical 
purposes, we place a more stringent bound on the core's motion, but only 
for a substantially shorter period of time.

Apart from its relevance for \GPB, our observations and astrometric
analysis are also of astrophysical
interest. The quasar 3C~454.3 is a highly active superluminal radio
source \citep[see, e.g.,][]{Pauliny-Toth+1987}.  It consists of a
relatively compact region from which a bent jet emanates
\citep[e.g.,][]{Pauliny-Toth1998}. Superluminal motion refers to those
apparent transverse velocities of the components within the source
that are measured to be greater than $c$, the speed of light. On the
basis of synchrotron radiation theory, the core is generally
identified as that component which is compact and has a flat or
inverted radio spectrum. Observations at 43 and 86 GHz show that
3C~454.3 has such a component located in the eastern part of the brightness
distribution \citep{Pagels+2004}.  Its characteristics are consistent
with those expected for the environment of a supermassive black hole
or the base of an associated jet, but not conclusively diagnostic of either one.
An additionally powerful probe for the location of a possible
supermassive black hole in an extragalactic source is to identify the
component in the radio structure of the source which shows the
smallest motion of all components or is stationary on the sky. Such a
component would be a strong candidate for being closely related to the
purported supermassive black hole which is likely to be both close to
the center of mass of the source and virtually stationary on the sky.

Placement of stringent limits on the proper motions of quasars and
other compact extragalactic radio sources distributed across the sky
are being made by others on a routine basis through astrometric/geodetic VLBI 
observations\footnote{Apart from source positions, these observations yield
antenna coordinates and velocities, and series of Earth orientation parameters.}.  
Random errors of frequently observed sources may be as low as 
6~$\mu$as, although systematic errors, mainly due to unaccounted
propagation effects and source structure are believed to be in the
range of 50--1000~$\mu$as.

All geodetic VLBI measurements are based on
interferometric group delays. More accurate measurements can be made, for example,
with targeted differential VLBI observations of two or more sources
located close to each other on the sky \citep{Shapiro+1979}.  In such
cases, interferometric phases or phase delays can be used, yielding
relative positions and proper motions of properly selected celestial
sources with uncertainties as low as $\sim$10~\muas~ and
$\sim$10~\muasyr, respectively \citep[e.g.,][]{MarcaideS1983,
Bartel+1986, RiojaP2000, BietenholzBR2001, FomalontK2003,
Brunthaler+2005}.

This paper is the third in a series of seven papers reporting on the
astrometric support for \GPB\ for the purpose of defining the
cosmological reference frame for the gyroscope precession
measurements.  In the first paper of this series we gave an overview
of the astronomical support for \GPB\ \citep[Paper I]{GPB-I}.  In the
second paper we focused on the characteristics of quasar 3C~454.3 and
the other two extragalactic reference sources, B2250+194 and
B2252+172, and reported on their structure and structure changes with
time and frequency \citep[Paper II]{GPB-II}. In this paper (Paper
III), we report on the degree of stationarity of the core of the
quasar 3C~454.3, which is the reference source for the guide star
\objectname[]{IM Peg} and therefore pivotal for \GPB\@. In the fourth
paper we present our VLBI astrometry analysis technique and compare it
with other such techniques \citep[Paper IV]{GPB-IV}.  In the fifth
paper we present our results for the proper motion and parallax of
\objectname[]{IM Peg} relative to the core of 3C~454.3 \citep[Paper
V]{GPB-V}. In the sixth paper we report on the orbital motion of
\objectname[]{IM Peg} and interpret the radio structure of the star
\citep[Paper VI]{GPB-VI}.  Finally, in the seventh paper, we focus on
the individual epochs of observation of \objectname[]{IM Peg} and
include a movie of the radio images of this star \citep[Paper
VII]{GPB-VII}.

Here we first briefly describe our observations, in \S~\ref{obs}.  We
give characteristics and show representative images of 3C~454.3,
B2250+194, and B2252+172 in \S~\ref{refsources}. We describe our
astrometry program in \S~\ref{astprog}.  We present astrometric
results in \S~\ref{ast1}, \ref{ast2}, and \ref{ast3}, discuss these
results in \S~\ref{discuss}, and give our conclusions in
\S~\ref{conclus}.

\section{Observations}
\label{obs}

As one of the strongest-emitting quasars at radio frequencies,
3C~454.3 has been observed in geodetic group-delay VLBI sessions since
1979.  For our \GPB\ VLBI program we made use of observations from the
total of 1119 such sessions between 1980 and 2008. In addition we used
geodetic observations of B2250+194 from a total of 38 sessions between
1996 and 2008 that were made in support of \GPB\ VLBI.

The bulk of our \GPB\ VLBI efforts were devoted to phase-delay VLBI
observations of \objectname[]{IM Peg} and our three reference sources,
3C~454.3, B2250+194, and B2252+172.  A detailed description of these
latter observations was given in Paper II; however, for the
convenience of the reader, we give a summary here.

We obtained 35 sets of 8.4 GHz VLBI observations in support of \GPB\
between 1997 January 16 and 2005 July 16.  We used a global array of
12 to 16 radio telescopes, which most often included MPIfR's 100 m
telescope at Effelsberg, Germany; NASA/Caltech/JPL's 70m DSN
telescopes at Robledo, Spain, Goldstone, CA, and Tidbinbilla,
Australia; NRAO's ten 25 m telescopes of the VLBA, across the U.S.A;
NRAO's phased VLA, equivalent to a 130 m telescope, near Socorro, NM;
and, at early times, NRCan's 46 m Algonquin Radio Telescope near
Pembroke, ONT, Canada, and, at later times, NRAO's 110 m GBT in
WV. In each session we made interleaved observations of 3C~454.3,
\objectname[]{IM Peg}, and B2250+194 by using a sequence of typically
3C~454.3 (80~s) - IM Peg (170~s) - 2250 (80~s).  For the last 12
sessions, starting 2002 November, we also observed B2252+172, but only
after every second sequence to allow greater concentration on the main
three sources.  The new observing sequence was 3C~454.3 (80~s) - IM
Peg (125~s) - 2250 (80~s) - 3C~454.3 (80~s) - IM Peg (125~s) - 2250
(80~s) - 2252 (90~s)\footnote{Here and hereafter we sometimes use as abbreviations
2250 for B2250+194 and 2252 for B2252+172.}. In three sessions we also observed at 5.0 and
15.0 GHz. All observations were recorded in both right and left
circular polarizations and processed on the VLBA hardware correlator at
Socorro.

\section{The Celestial Reference Sources}
\label{refsources}

\subsection{Sky Positions and Cosmological Distances}

\begin{figure}
\centering
\includegraphics[width=\textwidth]{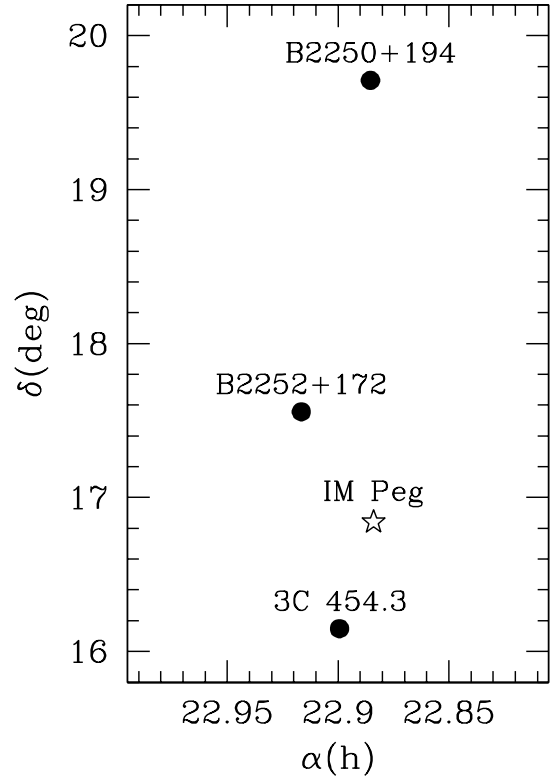} 
\caption{A sky chart with coordinates of the three reference sources
and the guide star, IM peg. The linear scale is the same for right
ascension ($\alpha$) and declination ($\delta$) for the declination of
3C~454.3.}
\label{f1skypos}
\end{figure}

In Figure~\ref{f1skypos} we show the positions of
\objectname[]{3C 454.3}, \objectname[]{B2250+194}, and
\objectname[]{B2252+172} along with that of the \GPB\ guide star,
\objectname[]{IM Peg}. All four sources are located approximately
along a single north-south axis allowing us to make easier use of
interpolation to estimate and potentially reduce the contributions of
the troposphere and the ionosphere to the total error in determining
the sources' relative positions. In Table~\ref{t1gen} we give the sky
separations of the two extragalactic sources from 3C~454.3, the
sources' flux densities, redshifts (when known), and angular diameter
distances, the latter assuming an inhomogeneous
Friedmann-Lema{\^i}tre-Robertson-Walker cosmology.
For comparison, we also give the characteristics of IM Peg.

\begin{deluxetable}{l l l l l l l}
\tabletypesize{\small}
\tablecaption{Characteristics of the sources}
\tablewidth{0pt}
\tablehead{
\colhead{Source} &  \colhead{Type} & \multicolumn{2}{c}{Separation}  & Flux density\tablenotemark{a} &   \colhead{Redshift}
                                    & \colhead{Distance}\tablenotemark{b} \\
                 &                 & \colhead{$\Delta$\al(\arcdeg)}  & \colhead{$\Delta$\de(\arcdeg)} &  \colhead{(Jy)} 
                 &                   & \colhead{(Mpc)} 
}
\startdata
3C~454.3 & quasar         & \nodata           & \nodata & 7  -- 10       & 0.859    & 1610       \\ 
B2250+194 & galaxy        &  $-0.2$           & 3.6     & 0.35   -- 0.45 & 0.28     & \phn880    \\
B2252+172 & unidentified  &     \phn 0.4      & 1.4     & 0.017          & \nodata  & \nodata     \\
\\
IM Peg  & RS CVn          &  $-0.1$           & 0.7     & 0.005 -- 0.05  & 0.0      & \phn\phn\phn0.0  \\
\enddata
\tablenotetext{a}{The range gives the lowest and highest flux density we measured at 8.4 GHz with the VLA during the course of our
observations, 1997 January to 2005 July.}
\tablenotetext{b}{The angular diameter distance for a flat universe with Hubble constant, 
$H_0$=70~\kmsMpc, and normalized density parameters, $\Omega_M=0.27$ and $\Omega_{\lambda}$=0.73 \citep{KantowskiKT2000}.} 
\label{t1gen}
\end{deluxetable}

\subsection{Representative Images}

\begin{figure}
\centering
\includegraphics[width=0.7\textwidth]{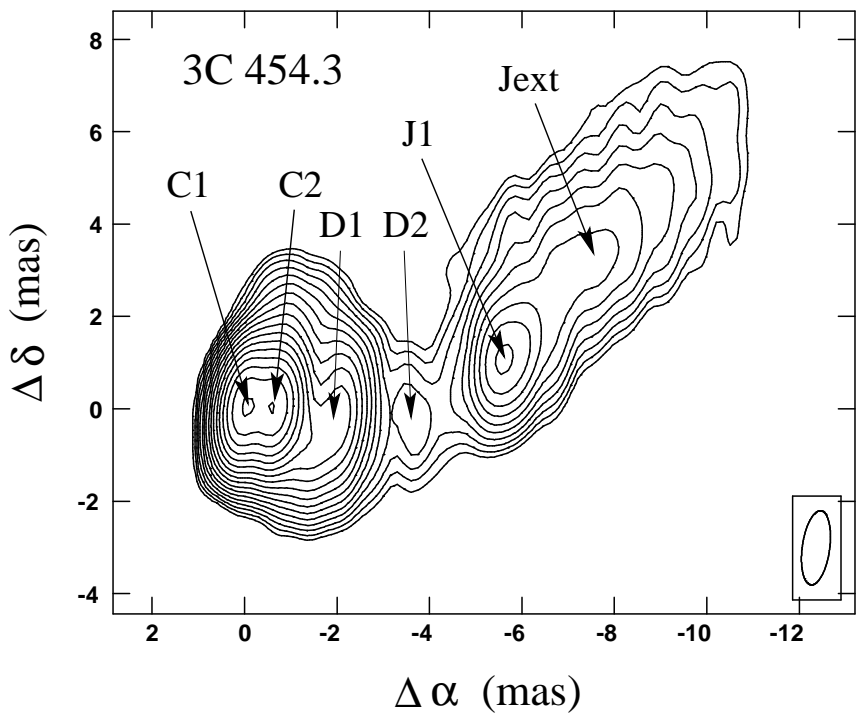}
\caption{An image of 3C~454.3 from observations on 2005 May 28 with
components C1, C2, D1, D2, J1, and Jext indicated. The contours start
at 10 m\Jb\ and increase by factors of $\sqrt{2}$ towards the
peak. The peak brightness is 2.68 \Jb. The rms brightness of the
background noise is 0.73 m\Jb. The full-width at half-maximum (FWHM)
contour of the Gaussian convolving beam is given in the lower
right. North is up and east to the left.}
\label{f23C454}
\end{figure}

\begin{figure}
\centering
\includegraphics[width=0.6\textwidth]{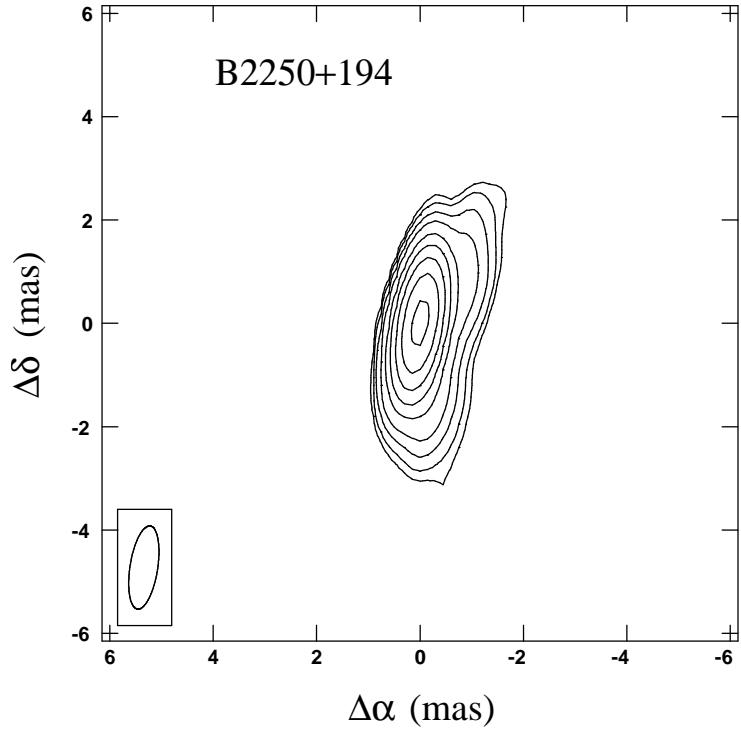}
\caption{An image of B2250+194 from observations on 2005 January 15.
The contours are at 0.3, 0.5, 1, 2, 5, 10, 20, 40, and 80\% of the
peak brightness of 0.43 \Jb. The rms brightness of the background noise in 0.08 \mJb.  The FWHM contour of the Gaussian
convolving beam is given in the lower left. North is up and east to
the left.}
\label{f32250}
\end{figure}

\begin{figure}
\centering
\includegraphics[width=0.6\textwidth]{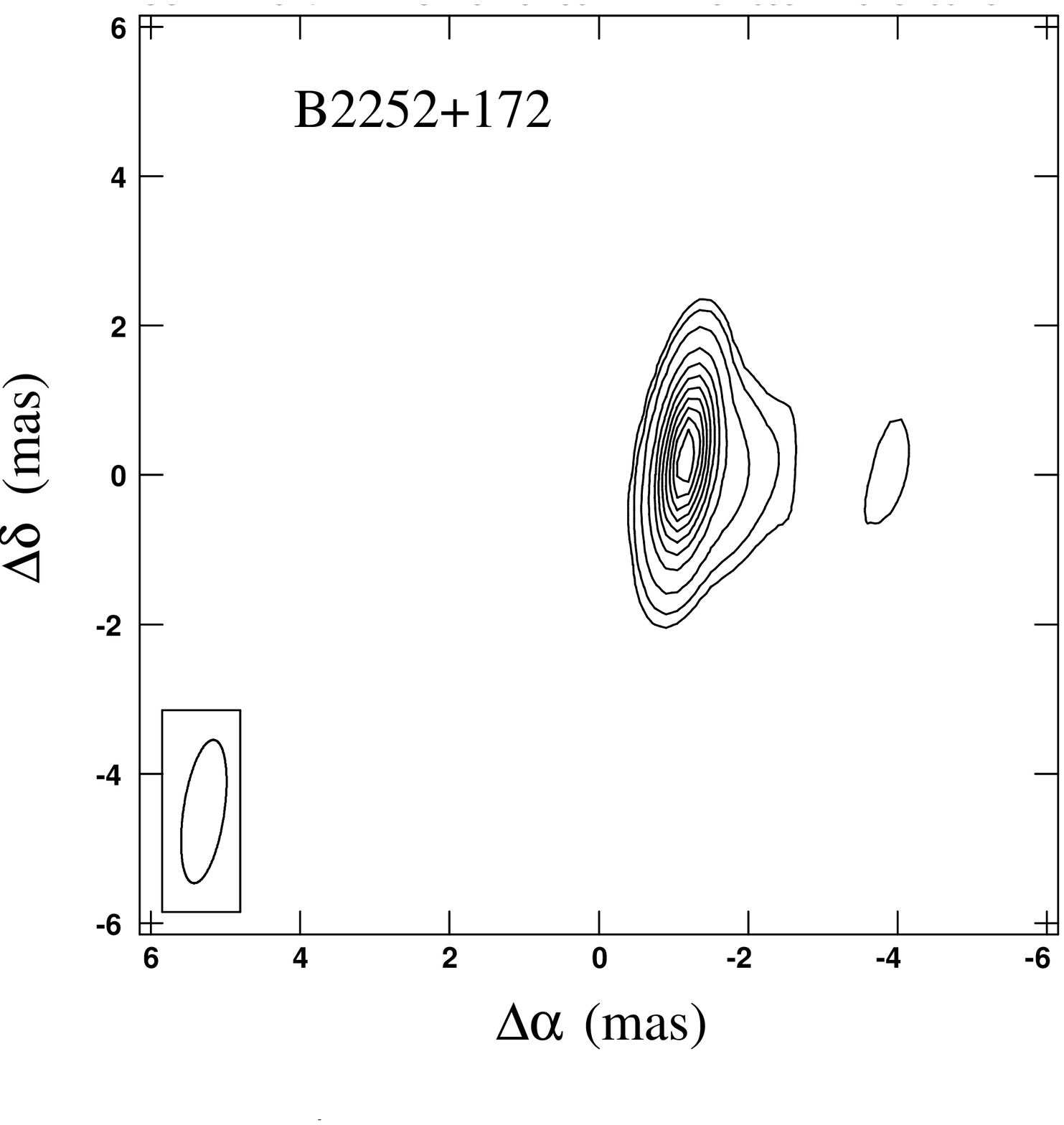}
\caption{An image of B2252+172 from observations on 2005 January 15.
The contours are at 3, 5, 10, 20, 30, \dots\ 90\% of the peak brightness
of 0.012 \Jb. The rms brightness of the background noise is 0.05 \mJb. 
The FWHM contour of the Gaussian convolving beam is
given in the lower left. North is up and east to the left.}
\label{f42252}
\end{figure}

In Figures~\ref{f23C454}, \ref{f32250}, and \ref{f42252} we show
representative images of \objectname[]{3C 454.3},
\objectname[]{B2250+194}, and \objectname[]{B2252+172}.  The source
3C~454.3 is a superluminal quasar with the highest radio flux density
of the three sources. It consists of a core region, which is primarily
extended east-west, and can be well modeled for each of our 35 epochs
by two compact components, C1 and C2, separated by $\sim$0.6 mas. When
studied at a higher resolution, e.g., at 43 and 86 GHz, the same
region has a complex structure \citep{GomezMA1999, Jorstad+2001b,
Jorstad+2005}, with C1 being essentially unresolved and having a size
at 86 GHz of $\leq$70~\muas\ \citep{Pagels+2004}. C2 is
extended. Also, between these two components there are others that
appear to move away from C1 toward C2 with superluminal speeds
\citep{Jorstad+2001b, Jorstad+2005}; these other components are not
individually visible in our lower-resolution 8.4 GHz images.

Further to the west there are components, D1 and D2, that are visible
only at later epochs in our observations. We modeled these as compact
components. They bridge the core region to the outer part of the jet
with a more than 10 mas extent in our images. This outer part of the
jet bends toward the northwest and in fact extends as far as a few
arcseconds away from the core \citep[see also][]{CawthorneG1996,
Pauliny-Toth1998}. The brightness peak of the extended 10 mas jet in
our image is clearly visible at each of our 35 epochs, and we modeled
it as a compact component which we call J1. The extended jet can be
modeled as an elliptical Gaussian, which we call Jext.

The source \objectname[]{B2250+194} is 20 times weaker in terms of
flux density than 3C~454.3, but much more compact, consisting of a
central component with north-south extensions and an apparent length of
$\sim$5~mas. The source \objectname[]{B2252+172} is the weakest of the
three in terms of flux density but also the most compact, consisting
of one dominant component and a barely visible extension to the
west. For more detail on the structure of these sources and their
evolution or lack thereof, see Paper II.

\subsection{Selection of 3C~454.3 as a Reference Source for \GPB}

Because of these differences in characteristics and separations from
\objectname[]{IM Peg}, each of these sources has advantages and
disadvantages as a reference source for \objectname[]{IM Peg}. The
quasar 3C~454.3 has the advantage of the highest flux density and
closest proximity to the guide star. The latter point is of chief
importance, since the standard errors of astrometric VLBI measurements
scale approximately linearly with the separation of the reference
source from the target source \citep{Shapiro+1979,
PradelCL2006}. Another advantage is that 3C~454.3 was used as the
astrometric reference source for \objectname[]{IM Peg} as early as
1991 \citep{Lestrade+1999}, thus extending the time baseline of VLBI position determinations and allowing increased accuracy in our proper-motion estimates. The main disadvantage of this reference source is its complex structure.

The advantage of the other two sources is their compact
structure. 
However, \objectname[]{B2250+194} is relatively far away on
the sky from \objectname[]{IM Peg}, and \objectname[]{B2252+172} has a
relatively low flux density. All things considered, we decided to use
3C~454.3 as the primary reference source for \objectname[]{IM Peg}.
Our task was then to find the component in 3C~454.3 that is most
closely associated with the dynamical center of the quasar and to test
the stationarity of this component with respect to our two other
reference sources and our CRF.

Images of 3C~454.3 at 8.4 and 15 GHz (Paper II) and at 43 and 86 GHz
\citep{Pagels+2004} show that the easternmost component, C1, remains
compact at the highest frequencies and angular resolutions yet investigated, and has a
flat or inverted spectrum in this frequency region. Other components
or condensations show structure at 86 GHz, and, in cases where it was
determinable, a steep spectral index. These characteristics indicate
that for our 8.4 GHz images, C1 is likely the component most closely
related to the  putative  supermassive  black hole and the quasar's  center of
mass.

\section{The Astrometry Program for \GPB: Goal, Strategy, and Procedure}
\label{astprog}

\begin{trivlist}
\item{\sl Goal: }Our main goal is to determine C1's position and
especially a bound on its proper motion relative to the distant
universe, to confirm its suitability as the primary reference for \GPB.

\item{\sl Strategy: }The distant universe is for our purposes most
usefully represented by compact, extragalactic radio sources. We choose
a procedure where we first determine the position and proper motion of C1
relative to our two reference sources, B2250+194 and B2252+172, and
second relative to our CRF.  For the first step,
interferometric phase-delays are used exclusively. This has the advantage of 
simplicity,  utilization of the same type of data (phase-delays)
and analysis technique, and highest precision and possibly highest
accuracy of the results. For the second, the
results from the first are added to position and proper-motion
determinations of B2250+194 in the CRF based on interferometric group
delays from the geodetic VLBI sessions. This has the advantage of having as a 
reference not only two very distant and compact sources, but $\sim$4000 sources
that define our CRF.   

\item{\sl Procedure:} We determine the position and the limit on proper motion of:

\begin{enumerate}
\item  C1 relative to
B2250+194 and B2252+172, and B2252+172 relative to B2250+194 in the
senses (C1 $-$ 2250), (C1 $-$ 2252), and (2252 $-$ 2250) using our
interleaved phase-delay VLBI observations. The combined result for C1
is obtained as a weighted mean of the first two
differenced solutions, with the third serving to demonstrate the 
consistency of our results and zero proper motion within the errors 
for the two sources relative to each other. 
Due to the source's compactness, any motions or brightness
distribution changes of these two reference sources
would likely be very small and therefore have only a marginal, if any, 
effect on our astrometric results for C1.

\item B2250+194 in the CRF using routine geodetic and astrometric group-delay VLBI
observations distributed by 
the International VLBI Service for Geodesy and Astrometry\footnote{Available 
at http://ivscc.gsfc.nasa.gov/products-data/index.html};

\item B2252+172 in the CRF
by adding the result from 2.\ to that from 1.\ in the sense (2252 $-$
2250) + 2250, and confirming the position result and proper motion limit by using recent single-epoch 
geodetic group-delay VLBI observations of B2252+172; and

\item C1 in the CRF in two
ways: First by adding the results from 1.\ to those from 2.: C1 = (C1
$-$ 2250) + 2250, and second by adding the results from 1. to those of
3.: C1 = (C1 $-$ 2252) + 2252.  The combined result is obtained as a
weighted mean from these two ways.

\end{enumerate}
\end{trivlist}

\section{Astrometric Results (1): Position Determinations for Each Observing Session}
\label{ast1}

\subsection{Analysis of Interleaved Phase-delay Observations of 3C 454.3, B2250+194, and B2252+172}

Our VLBI data for the reference sources, 3C~454.3, B2250+194, B2252+172,
and also from IM Peg, were analyzed
with an astrometric software package that was developed specifically
for the analysis of the \GPB\ VLBI data. It includes a
phase-connection program that automatically resolves 2$\pi$
ambiguities that exist in the set of VLBI phases for each
baseline so as to convert them to phase delays. The software also
includes in the phase-delay fitting \citep[e.g.,][]{Shapiro+1979,
Bartel+1986} a Kalman filter (see Paper IV) to model the variations of
the troposphere, the ionosphere, and the clock offsets at each VLBI
site \citep{Lebach+1999}.  In addition, we used two different models
to initially correct for the effects of the ionosphere, one we call
``JPL'' which is part of NRAO's imaging package, AIPS, and based on
GPS data provided by JPL, and the other, the older parametrized
ionospheric model ``PIM,'' developed at USAF Research Labs
\citep[described by, e.g.,][]{Campbell+1999}.

To relate the phase delays to a particular reference point in
3C~454.3, namely the core component C1, all phase delays from
differential astrometry involving 3C~454.3 were corrected for the
structure of 3C 454.3, as represented by the CLEAN components produced with AIPS;
C1 served as the phase reference
point.  The other two extragalactic sources were deemed sufficiently
compact for our purposes so that the reference point for each could be assumed
to be the brightness peak in its image.  We elaborate on this method
of astrometric VLBI data analysis and compare it to other methods in
Paper IV.

\subsection{Positions of the Components of 3C 454.3 Relative to Those of B2250+194 and B2252+172}

To test the positional stability of 3C~454.3's component C1 relative
to our two reference sources, we determined, for each of our 35
sessions of 8.4 GHz observations, the coordinates of C1, and for
comparison also those of C2, D1, D2, J1, and Jext, all relative to the
brightness peak of B2250+194 and, for the last 12 epochs, also to that of
B2252+172. We obtained two sets of coordinates by correcting for the
effects of the ionosphere in two different ways, one set by using the
JPL model and another set by using PIM\@.  Although the JPL model was
not available for our first 8 epochs but only from 1998 September 17
onward, we nevertheless in this paper use mainly phase-delay data corrected with the JPL model
since it proved to be superior in that it resulted in
smaller uncertainties of our astrometric estimates despite precluding
the use of our earliest phase-delay data. We elaborate on the comparison below. (In
Paper V we use PIM instead, because in that paper errors in modeling the 
ionosphere play a less significant role than they play here, and because
PIM has the advantage that it can be used for all of
our VLBI data.)
We list our coordinate determinations for C1, C2, D1, D2, J1, and Jext relative
to the brightness peak of \objectname[]{B2250+194} in
Table~\ref{t32250} and to that of \objectname[] {B2252+172} in
Table~\ref{t42252}, all obtained with the JPL model.

\subsection{The Uncertainties of the Relative Positions of the Components of 3C 454.3}
\label{errors} 

The uncertainties of the coordinates listed in Tables~\ref{t32250} and \ref{t42252} were
determined partly empirically, 
namely by adding a constant in quadrature to the statistical standard
errors so as to obtain a reduced Chi-square of unity ($\chinu=1$,
where $\nu$ is the number of degrees of freedom) in our residuals after solving for
relative position and proper motion in \al\ and \de\ separately.
This constant is assumed to approximately reflect
non-statistical errors.
Accurate standard errors are difficult to estimate in any other way.  They contain
contributions from noise and from systematic errors, with the latter
due mostly to deconvolution, source structure, and atmospheric and
ionospheric variations.  We next discuss and approximately quantify each contribution in turn. 

\subsubsection{Noise}

Noise in an image has a number of sources.  The rms background brightness
in the images is dominated by contributions from statistical noise in
the radio signals and thermal noise in the receivers.  
However, for our relatively high
dynamic range images (typically over 2,000 to 1; see,
e.g., Figures~\ref{f23C454} to \ref{f42252}), the various uncertainties in the bright
parts of the image are larger than their corresponding rms background brightnesses
 and are dominated by contributions which are not
strictly random such as residual calibration errors and deconvolution
errors.
Given the small rms of the background brightness relative to the
peaks in the maps, we conclude that this noise causes errors in the estimate of the 
separation of each of components C1 and C2 in 3C~454.3 from B2250+194
by a correspondingly small portion of the HWHM (half-width at 
half-maximum) of the beam, namely by $<$5~\muas, and of each of components D1,
D2, J1 and Jext in 3C~454.3 from B2250+194 by $<$10~\muas.  The
corresponding errors in our estimates of the separation of these components from the
weaker source, B2252+172, are dominated by the source's lower
peak-to-noise ratio but are still $<$10~\muas.

\subsubsection{Deconvolution errors} 

Deconvolution errors are caused by the visibility measurements not
filling the \uv~plane of the VLBI array up to its highest angular
resolution and by the resulting generation of side lobes in the image
plane, which are not completely eliminated through the deconvolution
process. 
We studied
this type of error by using a noise-free model image similar to the
image of 3C~454.3 at 8.4 GHz, 
Fourier-transforming the model to the \uv\ plane, and then using the
same \uv-plane sampling as in one of our typical observing sessions.
The
generated \uv\ model data were then used for imaging and deconvolution. We then
determined the difference between the position of each component in the
model image and the position of the corresponding component in the
deconvolved image. We found that the deconvolution
error for each coordinate of each of the six components of 3C~454.3 is
typically $30$~\muas\ and never larger than $\sim40$~\muas.  The
deconvolution errors for sources like B2250+194 and B2252+172 with
relatively simple brightness distributions are doubtless smaller given
the same \uv\ coverage as for 3C~454.3. We therefore conclude that the
standard error in each coordinate of the separation of any component
in 3C~454.3 from either B2250+194 or B2252+172 is typically
$30$~\muas.

\subsubsection{Structure errors} 

Structure errors are caused by a mismatch between the Gaussian
component model and the brightness distribution of the source. If the
source were completely unresolved, a fit of a Gaussian to the image
with the parameters of the convolving beam would give the position of
the source with an essentially zero structure error. Indeed, the
sources B2250+194 and B2252+172 are rather compact, with brightness
peaks that can be clearly identified and located, and hence we expect the structure
errors to be small in comparison to those for the
components in 3C~454.3. As can be seen in Figure~\ref{f23C454},
components C1 and C2 are located close together in comparison to the
size of the beam and can barely be distinguished at several epochs
(see Paper II).  In addition, there is ``confusing'' emission in the
neighborhood of these two components. These two characteristics cause
significant structure errors, difficult to estimate quantitatively,
but likely as large as a good fraction of the HWHM of the beam.  The
components D1 and D2 are weaker than C1 and C2, and visible only at later
epochs, but then clearly distinguishable.  Their identification is
likely as uncertain as that of  C1 and C2.  Component J1 is
always visible as a single peak; however, it is embedded in the
extended brightness distribution of the 10 mas long jet rendering the
identification of a component in that jet as uncertain as, for many
epochs, the identification of C1 and C2.  The identification of Jext is
more uncertain than that of any of the other components since this component
represents the extended brightness distribution that J1 is embedded
in. The extension is largely toward the northwest, thus affecting both
position coordinates of Jext. A detailed analysis of these uncertainties 
is difficult to carry out quantitatively, in view of the unknown 
characteristics of the relevant structures. Based on our experience, 
however, we expect structure errors in each
of the position coordinates to be $\sim30$~\muas\ for components C1, C2,
D1, D2, and J1 and $\sim50$~\muas\ for the component Jext. In comparison, the
structure errors for B2250+194 and B2252+172 are negligible.

\subsubsection{Residual propagation-medium errors}

Using our Kalman filter \citep[Paper IV]{GPB-IV} removes a large portion of the
propagation-medium errors from the estimates of separation between two
sources. Nevertheless, errors in modeling the troposphere and the
ionosphere at each site likely still represent the largest sources of
error in our estimates of relative positions. These errors scale
approximately with the separation between the sources
\citep{Shapiro+1979, PradelCL2006}.  These errors also depend on the model
used for the ionosphere (JPL or PIM).

\begin{figure}
\centering
\includegraphics[width=\textwidth]{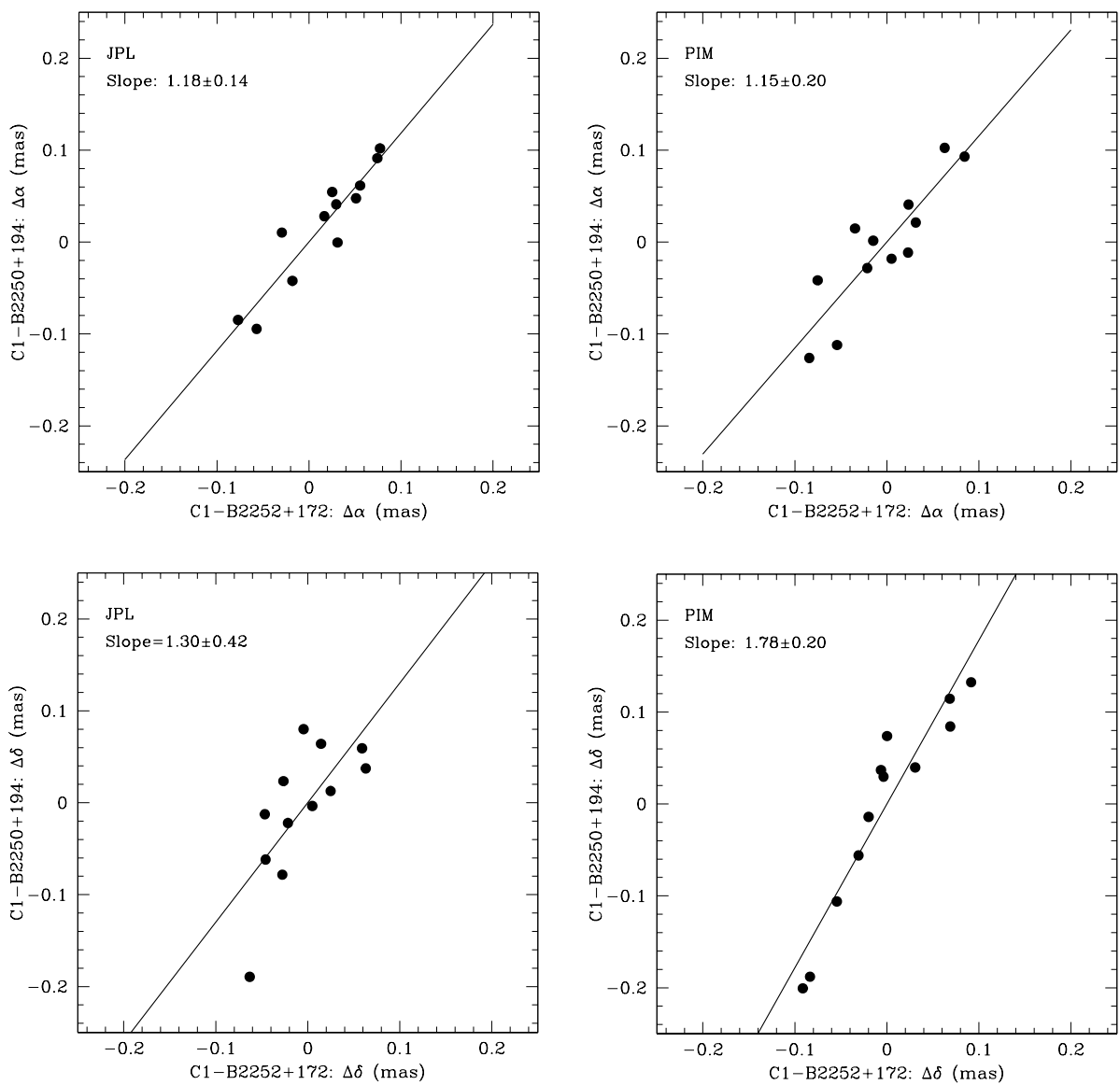}
\caption{The coordinate differences (except for offsets) in \al\
(upper row) and \de\ (lower row) of B2250+194 $-$ C1 as a function of
the coordinate differences of B2252+172 $-$ C1 for data for which the
ionospheric effects were corrected with the JPL model (left column)
and with PIM (right column).  The data corrected with the JPL model
were taken from Tables~\ref{t32250} and \ref{t42252}. The data
corrected with PIM are not given in the tables. The straight lines
indicate the least-squares fits. The slope and its statistical
standard error are given for the fit in each figure.}
\label{f7lcor}
\end{figure}

In our case the sources are oriented approximately north-south with the
separations in \al\ of B2250+194 and B2252+172 from C1 being about
equal, but with the separation in \de\ of B2252+172 from C1 being only
about 40\% of that for B2250+194 from C1. Also, the position
determinations are more affected in \de\ due to the propagation
medium's distorting effects depending strongly on the elevation angle of the source.
Therefore, any residual uncorrected ionospheric or tropospheric
effects should be most visible in comparisons of the estimated declinations
of C1 relative to B2250+194 with those relative to
B2252+172, since the angular separation of B2250+194 and B2252+172 is
predominantly north-south.  To evaluate the adequacy of the models, we plot
the relative coordinates, C1 $-$ B2250+194, as a function of the
relative coordinates, C1 $-$ B2252+172, separately in \al\ and
\de. For instance, a straight-line slope significantly larger than
unity would indicate uncorrected ionospheric or tropospheric
effects. We plot the relative coordinates for the JPL model and for
PIM in Figure~\ref{f7lcor}. The data corrected with the JPL model were
taken from Tables~\ref{t32250} and \ref{t42252}, whereas the data
corrected with PIM are not listed.

Least-squares fits\footnote{For simplicity we fit the data via
least-squares only in the vertical direction. Fitting the data in both
directions would not significantly alter our results. Also, since the
errors in the vertical direction are similar to each other and the
errors in the horizontal direction are similar to each other too, we
did not weight the data in the fits and also did not plot the errors in the
figures.} to the data give slopes that are all larger than
unity, reflecting the larger separation of B2250+194 from C1 than of
B2252+172 from C1.  
However, the slopes are within 1.3$\sigma$ of unity, except for the
data in \de\ for the PIM model. Here the slope is $1.78\pm0.20$. It is
smaller than a slope of $\sim$2.5 which, if there were no ionospheric
and tropospheric correction, would likely be expected on the basis of
the $\sim$2.5:1 ratio of the separations of the two sources from
C1. However, this slope is significantly ($\sim4\sigma$) larger than unity and also larger
than the corresponding slope for the data corrected with the JPL
model.  The JPL model thus seems to provide a better correction for the
ionosphere.

An approximate estimate of uncorrected errors due to the ionosphere
and troposphere for position estimates using PIM (as in Paper V for IM
Peg) and an upper limit of such errors for position estimates using
the JPL model (this paper) can be obtained from a close inspection of
the declination data corrected with PIM (lower right panel of
Figure~\ref{f7lcor}).  While the $\Delta$\de\ values of C1 $-$
2252 vary over a range of 0.19 mas, the corresponding values of
C1 $-$ 2250 vary over a range of 0.34 mas, 1.78 times larger, as
also indicated by the slope. Therefore, the difference of the
variations of 0.15 mas can likely be attributed to effects of the
ionosphere left uncorrected by PIM. Consequently, we think that for
data corrected with PIM the peak-to-peak contribution to the error of
the position difference, C1 $-$ 2250, in \de\ is approximately
$\pm$0.08~mas. To be conservative, we take this as a standard error.
The corresponding contribution to the error of the position
difference, C1 $-$ 2252, in \de\ approximately scales with the
separation and is therefore likely to be about 0.03 mas. The
contributions to the errors in \al\ that we estimated from scaling are
0.04 mas and again 0.03 mas for the two position differences,
respectively. For data corrected with the JPL model, we considered all these
estimates for PIM as still more conservative upper limits for the
standard deviations of these errors.

\subsubsection{Total errors}

Adding in quadrature the errors from noise, deconvolution, structure,
and propagation, we obtain estimated standard errors in \al\ of the
position differences from B2250+194 for the components C1, C2, D1, D2,
and J1 of 0.06 mas and for the component Jext of 0.07 mas. The
corresponding standard errors in \de\ are 0.09 mas and 0.10 mas.  The
corresponding standard errors of the position differences from
B2252+172, in both \al\ and \de, for the components C1, C2, D1, D2,
and J1 are 0.05 mas and for the component Jext 0.07 mas.

\subsubsection{Comparison with empirically determined errors}

For the position differences of these components from B2250+194, our
estimated standard errors are somewhat smaller than the empirically
determined standard errors which were mostly between the values of
0.08 and 0.10 mas in both \al\ and \de\ for these components
(see Table \ref{t32250}).  The reason is the large variation of some
of the data points near epoch 2001
which increased the empirically determined errors.  For the
corresponding position differences from B2252+172, all for epochs
after 2001, these estimated standard errors agree well with the
empirically determined standard errors of 0.04 to 0.06 mas in \al\ and
\de, respectively, for the components C1, C2, D1, D2, and J1 and of
0.07 mas in \al\ and \de\ for the component Jext (see Table
\ref{t42252}).

\begin{deluxetable}{r c r r r r r r r r r r r r}
\rotate
\tabletypesize{\tiny}
\tablecaption{Separation of 3C 454.3 components from B2250+194}
\tablewidth{0pt}
\tablehead{
\colhead{Epoch} & \colhead{Julian date} & \multicolumn{2}{c}{C1\tablenotemark{a}}  & \multicolumn{2}{c}{C2\tablenotemark{a}}  & \multicolumn{2}{c}{D1\tablenotemark{a}}  & \multicolumn{2}{c}{D2\tablenotemark{a}}  & \multicolumn{2}{c}{J1\tablenotemark{a}} & \multicolumn{2}{c}{Jext\tablenotemark{a}} \\
 &  \colhead{2450000.0 +}   & \colhead{$\Delta$\al}  & \colhead{$\Delta$\de}
                 & \colhead{$\Delta$\al}  & \colhead{$\Delta$\de} 
                 & \colhead{$\Delta$\al}  & \colhead{$\Delta$\de}
                 & \colhead{$\Delta$\al}  & \colhead{$\Delta$\de}         
                 & \colhead{$\Delta$\al}  & \colhead{$\Delta$\de}         
                 & \colhead{$\Delta$\al}  & \colhead{$\Delta$\de}                     \\
               &  & \colhead{(mas)}               & \colhead{(mas)}
                 & \colhead{(mas)}               & \colhead{(mas)}
                 & \colhead{(mas)}               & \colhead{(mas)}       
                 & \colhead{(mas)}               & \colhead{(mas)}        
                 & \colhead{(mas)}               & \colhead{(mas)}           
                 & \colhead{(mas)}               & \colhead{(mas)}  
}            
\startdata
1998 Sep 17 & 1073.8     &  -0.11$\pm$0.08 &   0.25$\pm$0.09 &  -0.58$\pm$0.11 &   0.22$\pm$0.10 & \nodata & \nodata & \nodata & \nodata &  -5.57$\pm$0.07 &   0.72$\pm$0.10 &  -6.34$\pm$0.10 &   2.09$\pm$0.08  \\
1999 Mar 13 & 1251.3     &  -0.21$\pm$0.08 &   0.11$\pm$0.08 &  -0.68$\pm$0.12 &   0.06$\pm$0.09 & \nodata & \nodata & \nodata & \nodata &  -5.63$\pm$0.07 &   0.61$\pm$0.09 &  -6.46$\pm$0.11 &   1.99$\pm$0.08  \\
1999 May 15 & 1314.1     &  -0.11$\pm$0.08 &   0.14$\pm$0.08 &  -0.52$\pm$0.12 &   0.07$\pm$0.09 & \nodata & \nodata & \nodata & \nodata &  -5.48$\pm$0.07 &   0.64$\pm$0.09 &  -6.29$\pm$0.11 &   1.98$\pm$0.08  \\
1999 Sep 18 & 1440.8     &  -0.27$\pm$0.08 &   0.23$\pm$0.08 &  -0.68$\pm$0.12 &   0.14$\pm$0.09 & \nodata & \nodata & \nodata & \nodata &  -5.66$\pm$0.07 &   0.74$\pm$0.10 &  -6.54$\pm$0.11 &   2.14$\pm$0.08  \\
1999 Dec 09 & 1522.5     &  -0.12$\pm$0.08 &   0.21$\pm$0.08 &  -0.60$\pm$0.12 &   0.16$\pm$0.09 & \nodata & \nodata & \nodata & \nodata &  -5.59$\pm$0.07 &   0.76$\pm$0.10 &  -6.52$\pm$0.10 &   2.24$\pm$0.08  \\
2000 May 15 & 1680.1     &  -0.10$\pm$0.08 &   0.16$\pm$0.08 &  -0.71$\pm$0.12 &   0.17$\pm$0.09 & \nodata & \nodata & \nodata & \nodata &  -5.64$\pm$0.08 &   0.87$\pm$0.08 &  -6.50$\pm$0.10 &   2.05$\pm$0.08  \\
2000 Aug 07 & 1763.9     &  -0.13$\pm$0.09 &   0.10$\pm$0.09 &  -0.78$\pm$0.14 &   0.08$\pm$0.10 & \nodata & \nodata & \nodata & \nodata &  -5.64$\pm$0.09 &   0.82$\pm$0.10 &  -6.59$\pm$0.11 &   2.17$\pm$0.09  \\
2000 Nov 05 & 1854.6     &  -0.26$\pm$0.08 &   0.06$\pm$0.08 &  -0.84$\pm$0.13 &  -0.01$\pm$0.09 & \nodata & \nodata & \nodata & \nodata &  -5.68$\pm$0.08 &   0.77$\pm$0.09 &  -6.53$\pm$0.10 &   1.99$\pm$0.08  \\
2000 Nov 06 & 1855.6     &  -0.44$\pm$0.13 &  -0.36$\pm$0.18 &  -1.14$\pm$0.16 &  -0.42$\pm$0.18 & \nodata & \nodata & \nodata & \nodata &  -5.93$\pm$0.12 &   0.35$\pm$0.19 &  -7.01$\pm$0.14 &   2.01$\pm$0.18  \\
2001 Mar 31 & 2000.2     &   0.06$\pm$0.09 &   0.19$\pm$0.10 &  -0.37$\pm$0.12 &   0.13$\pm$0.11 & \nodata & \nodata & \nodata & \nodata &  -5.42$\pm$0.08 &   0.96$\pm$0.10 &  -6.48$\pm$0.11 &   2.34$\pm$0.09  \\
2001 Jun 29 & 2090.0     &  -0.05$\pm$0.08 &   0.14$\pm$0.08 &  -0.59$\pm$0.14 &   0.09$\pm$0.09 & \nodata & \nodata & \nodata & \nodata &  -5.57$\pm$0.09 &   0.95$\pm$0.09 &  -6.50$\pm$0.11 &   2.22$\pm$0.08  \\
2001 Oct 19 & 2202.7     &  -0.12$\pm$0.11 &   0.02$\pm$0.15 &  -0.70$\pm$0.14 &  -0.05$\pm$0.16 & \nodata & \nodata & \nodata & \nodata &  -5.65$\pm$0.11 &   0.88$\pm$0.16 &  -6.68$\pm$0.13 &   2.29$\pm$0.15  \\
2001 Dec 21 & 2265.5     &  -0.21$\pm$0.08 &  -0.06$\pm$0.08 &  -0.89$\pm$0.14 &  -0.13$\pm$0.09 & \nodata & \nodata & \nodata & \nodata &  -5.75$\pm$0.08 &   0.79$\pm$0.09 &  -6.81$\pm$0.11 &   2.28$\pm$0.08  \\
2002 Apr 14 & 2379.2     &  -0.09$\pm$0.09 &   0.12$\pm$0.11 &  -0.64$\pm$0.12 &   0.01$\pm$0.12 & \nodata & \nodata & \nodata & \nodata &  -5.58$\pm$0.09 &   1.06$\pm$0.13 &  -6.65$\pm$0.11 &   2.41$\pm$0.11  \\
2002 Jul 14 & 2469.9     &  -0.09$\pm$0.08 &   0.10$\pm$0.09 &  -0.60$\pm$0.14 &   0.00$\pm$0.10 & \nodata & \nodata & \nodata & \nodata &  -5.59$\pm$0.08 &   0.98$\pm$0.09 &  -6.65$\pm$0.11 &   2.34$\pm$0.08  \\
2002 Nov 20 & 2599.6     &  -0.12$\pm$0.09 &  -0.16$\pm$0.10 &  -0.69$\pm$0.13 &  -0.25$\pm$0.11 & \nodata & \nodata & \nodata & \nodata &  -5.71$\pm$0.09 &   0.77$\pm$0.11 &  -6.79$\pm$0.11 &   2.22$\pm$0.09  \\
2003 Jan 26 & 2666.4     &  -0.08$\pm$0.09 &   0.06$\pm$0.10 &  -0.67$\pm$0.13 &  -0.06$\pm$0.11 & -1.56$\pm$0.07 &  -0.20$\pm$0.07 & \nodata & \nodata & -5.65$\pm$0.09 &   0.95$\pm$0.11 &  -6.67$\pm$0.11 &   2.28$\pm$0.09  \\
2003 May 18 & 2778.1     &   0.05$\pm$0.08 &   0.05$\pm$0.08 &  -0.64$\pm$0.12 &  -0.01$\pm$0.09 & -1.51$\pm$0.08 &  -0.24$\pm$0.06 & \nodata & \nodata & -5.64$\pm$0.08 &   1.03$\pm$0.10 &  -6.59$\pm$0.10 &   2.30$\pm$0.08  \\
2003 Sep 08 & 2891.7     &   0.02$\pm$0.08 &   0.02$\pm$0.09 &  -0.69$\pm$0.12 &  -0.06$\pm$0.10 & -1.62$\pm$0.08 &  -0.30$\pm$0.07 & \nodata & \nodata & -5.60$\pm$0.11 &   1.00$\pm$0.11 &  -6.61$\pm$0.11 &   2.28$\pm$0.09  \\
2003 Dec 05 & 2979.5     &  -0.01$\pm$0.08 &   0.11$\pm$0.09 &  -0.73$\pm$0.11 &   0.01$\pm$0.10 & -1.70$\pm$0.09 &  -0.27$\pm$0.06 & \nodata & \nodata & -5.59$\pm$0.08 &   1.03$\pm$0.10 &  -6.63$\pm$0.10 &   2.37$\pm$0.08  \\
2004 Mar 06 & 3071.3     &   0.02$\pm$0.08 &   0.00$\pm$0.09 &  -0.68$\pm$0.12 &  -0.10$\pm$0.10 & -1.66$\pm$0.09 &  -0.37$\pm$0.05 & -3.53$\pm$0.05 &  -0.50$\pm$0.13 &  -5.56$\pm$0.08 &   0.96$\pm$0.10 &  -6.62$\pm$0.10 &   2.38$\pm$0.08  \\
2004 May 18 & 3144.1     &  -0.03$\pm$0.09 &   0.09$\pm$0.09 &  -0.71$\pm$0.12 &  -0.06$\pm$0.10 & -1.69$\pm$0.07 &  -0.29$\pm$0.07 & -3.59$\pm$0.05 &  -0.41$\pm$0.12 &  -5.57$\pm$0.10 &   1.05$\pm$0.12 &  -6.72$\pm$0.11 &   2.53$\pm$0.09  \\
2004 Jun 26 & 3183.0     &  -0.13$\pm$0.08 &  -0.04$\pm$0.09 &  -0.83$\pm$0.16 &  -0.18$\pm$0.10 & -1.86$\pm$0.07 &  -0.37$\pm$0.06 & -3.69$\pm$0.05 &  -0.53$\pm$0.10 &  -5.61$\pm$0.10 &   0.91$\pm$0.10 &  -6.73$\pm$0.11 &   2.31$\pm$0.08  \\
2004 Dec 11 & 3351.5     &   0.01$\pm$0.08 &   0.01$\pm$0.08 &  -0.67$\pm$0.12 &  -0.06$\pm$0.09 & -1.66$\pm$0.10 &  -0.44$\pm$0.05 & -3.70$\pm$0.04 &  -0.39$\pm$0.10 &  -5.62$\pm$0.09 &   0.94$\pm$0.09 &  -6.77$\pm$0.10 &   2.45$\pm$0.07  \\
2005 Jan 15 & 3386.4     &   0.06$\pm$0.08 &  -0.05$\pm$0.08 &  -0.62$\pm$0.12 &  -0.10$\pm$0.09 & -1.66$\pm$0.10 &  -0.48$\pm$0.05 & -3.64$\pm$0.05 &  -0.55$\pm$0.10 &  -5.57$\pm$0.08 &   0.88$\pm$0.09 &  -6.75$\pm$0.10 &   2.42$\pm$0.08  \\
2005 May 28 & 3519.1     &   0.00$\pm$0.08 &   0.04$\pm$0.08 &  -0.69$\pm$0.12 &   0.00$\pm$0.09 & -1.82$\pm$0.09 &  -0.40$\pm$0.05 & -3.71$\pm$0.05 &  -0.23$\pm$0.10 &  -5.63$\pm$0.08 &   1.02$\pm$0.10 &  -6.75$\pm$0.10 &   2.52$\pm$0.08  \\
2005 Jul 16 & 3567.9     &  -0.04$\pm$0.08 &   0.09$\pm$0.09 &  -0.76$\pm$0.12 &   0.06$\pm$0.10 & -1.92$\pm$0.09 &  -0.34$\pm$0.06 & -3.74$\pm$0.05 &  -0.23$\pm$0.11 &  -5.69$\pm$0.08 &   1.05$\pm$0.11 &  -6.86$\pm$0.10 &   2.60$\pm$0.08  \\

\enddata \tablenotetext{a}{The coordinate differences of the
components of 3C 454.3 from those of B2250+194 (3C~454.3 - 2250)
for each epoch for which the JPL model for the correction of the
ionospheric effects could be used: \Ra{00}{00}{50}{3787837} +
$\Delta$\al\ and \dec{-3}{33}{41}{067505} + $\Delta$\de. The
coordinate differences are based on our differential measurements of
C1 relative to B2250+194 and on the determinations of C2, D1, D2, J1,
and Jext relative to C1 (Paper II). For B2250+194 the CRF coordinates
\Ra{22}{53}{7}{3691736} and \dec{19}{42}{34}{628786} (solution
\#3, Table~\ref{t2icrf}) were used. The standard errors are the
statistical standard errors with a constant added in
quadrature so that $\chinu = 1.$}
\label{t32250}
\end{deluxetable}

\begin{deluxetable}{r@{}c@{}c c c c c c c c c c c c}
\rotate
\tabletypesize{\tiny}
\tablecaption{Separation of 3C 454.3 components from B2252+172}
\tablewidth{0pt}
\tablehead{
\colhead{Epoch} & \colhead{Julian} & \multicolumn{2}{c}{C1\tablenotemark{a}}  & \multicolumn{2}{c}{C2\tablenotemark{a}}  & \multicolumn{2}{c}{D1\tablenotemark{a}}  & \multicolumn{2}{c}{D2\tablenotemark{a}}  & \multicolumn{2}{c}{J1\tablenotemark{a}} & \multicolumn{2}{c}{Jext\tablenotemark{a}} \\
                & \colhead{date} \\
 &  \colhead{2450000+}   & \colhead{$\Delta$\al}  & \colhead{$\Delta$\de}
                 & \colhead{$\Delta$\al}  & \colhead{$\Delta$\de} 
                 & \colhead{$\Delta$\al}  & \colhead{$\Delta$\de}
                 & \colhead{$\Delta$\al}  & \colhead{$\Delta$\de}         
                 & \colhead{$\Delta$\al}  & \colhead{$\Delta$\de}                     \\
               &  & \colhead{(mas)}               & \colhead{(mas)}
                 & \colhead{(mas)}               & \colhead{(mas)}
                 & \colhead{(mas)}               & \colhead{(mas)}       
                 & \colhead{(mas)}               & \colhead{(mas)}           
                 & \colhead{(mas)}               & \colhead{(mas)}    
}            

\startdata
2002 Nov 20 & 2599.6   &  -0.11$\pm$0.05 &  -0.06$\pm$0.04 &  -0.68$\pm$0.06 &  -0.15$\pm$0.05 & \nodata        & \nodata         & \nodata        & \nodata         & -5.70$\pm$0.05 &   0.87$\pm$0.04 &  -6.78$\pm$0.07 &   2.32$\pm$0.07  \\
2003 Jan 26 & 2666.4   &  -0.05$\pm$0.05 &   0.07$\pm$0.04 &  -0.64$\pm$0.07 &  -0.06$\pm$0.05 & -1.53$\pm$0.05 &  -0.20$\pm$0.06 & \nodata        & \nodata         & -5.62$\pm$0.06 &   0.95$\pm$0.05 &  -6.64$\pm$0.07 &   2.28$\pm$0.07  \\
2003 May 18 & 2778.1   &   0.04$\pm$0.05 &  -0.02$\pm$0.04 &  -0.65$\pm$0.03 &  -0.09$\pm$0.04 & -1.53$\pm$0.07 &  -0.32$\pm$0.05 & \nodata        & \nodata         & -5.65$\pm$0.01 &   0.96$\pm$0.05 &  -6.61$\pm$0.06 &   2.23$\pm$0.06  \\
2003 Sep 08 & 2891.7   &   0.02$\pm$0.05 &   0.01$\pm$0.04 &  -0.69$\pm$0.04 &  -0.08$\pm$0.04 & -1.62$\pm$0.06 &  -0.32$\pm$0.05 & \nodata        & \nodata         & -5.61$\pm$0.08 &   0.99$\pm$0.05 &  -6.61$\pm$0.06 &   2.27$\pm$0.07  \\
2003 Dec 05 & 2979.5   &  -0.02$\pm$0.05 &   0.00$\pm$0.04 &  -0.74$\pm$0.03 &  -0.10$\pm$0.05 & -1.71$\pm$0.07 &  -0.37$\pm$0.05 & \nodata        & \nodata         & -5.60$\pm$0.04 &   0.93$\pm$0.05 &  -6.64$\pm$0.06 &   2.27$\pm$0.07  \\
2004 Mar 06 & 3071.3   &  -0.01$\pm$0.05 &  -0.02$\pm$0.04 &  -0.70$\pm$0.03 &  -0.12$\pm$0.05 & -1.69$\pm$0.07 &  -0.39$\pm$0.05 & -3.56$\pm$0.03 &  -0.52$\pm$0.11 & -5.59$\pm$0.03 &   0.93$\pm$0.05 &  -6.64$\pm$0.06 &   2.36$\pm$0.07  \\
2004 May 18 & 3144.1   &  -0.06$\pm$0.05 &   0.06$\pm$0.05 &  -0.74$\pm$0.05 &  -0.08$\pm$0.05 & -1.73$\pm$0.06 &  -0.32$\pm$0.05 & -3.62$\pm$0.03 &  -0.43$\pm$0.10 & -5.61$\pm$0.07 &   1.03$\pm$0.07 &  -6.76$\pm$0.06 &   2.50$\pm$0.07  \\
2004 Jun 26 & 3183.0   &  -0.09$\pm$0.05 &  -0.04$\pm$0.04 &  -0.79$\pm$0.11 &  -0.18$\pm$0.05 & -1.82$\pm$0.05 &  -0.38$\pm$0.05 & -3.65$\pm$0.02 &  -0.53$\pm$0.08 & -5.57$\pm$0.06 &   0.90$\pm$0.04 &  -6.69$\pm$0.06 &   2.30$\pm$0.06  \\
2004 Dec 11 & 3351.5   &   0.02$\pm$0.05 &  -0.04$\pm$0.04 &  -0.66$\pm$0.04 &  -0.11$\pm$0.04 & -1.65$\pm$0.08 &  -0.49$\pm$0.05 & -3.69$\pm$0.02 &  -0.45$\pm$0.08 & -5.61$\pm$0.05 &   0.89$\pm$0.05 &  -6.76$\pm$0.06 &   2.40$\pm$0.06  \\
2005 Jan 15 & 3386.4   &   0.04$\pm$0.05 &  -0.02$\pm$0.04 &  -0.64$\pm$0.04 &  -0.08$\pm$0.04 & -1.69$\pm$0.09 &  -0.46$\pm$0.05 & -3.66$\pm$0.03 &  -0.52$\pm$0.08 & -5.59$\pm$0.04 &   0.91$\pm$0.04 &  -6.77$\pm$0.06 &   2.45$\pm$0.06  \\
2005 May 28 & 3519.1   &   0.00$\pm$0.05 &   0.03$\pm$0.04 &  -0.70$\pm$0.03 &  -0.01$\pm$0.04 & -1.83$\pm$0.08 &  -0.41$\pm$0.05 & -3.72$\pm$0.03 &  -0.24$\pm$0.08 & -5.64$\pm$0.03 &   1.01$\pm$0.04 &  -6.76$\pm$0.06 &   2.51$\pm$0.06  \\
2005 Jul 16 & 3567.9   &   0.00$\pm$0.05 &   0.02$\pm$0.04 &  -0.73$\pm$0.03 &  -0.01$\pm$0.04 & -1.89$\pm$0.07 &  -0.41$\pm$0.05 & -3.71$\pm$0.02 &  -0.30$\pm$0.09 & -5.66$\pm$0.03 &   0.97$\pm$0.06 &  -6.82$\pm$0.06 &   2.52$\pm$0.06  \\
\enddata

\tablenotetext{a} {As in Table~\ref{t32250} but now with B2252+172 as a reference. 
The coordinate differences (3C454.3 $-$ 2252) are \Ra{-00}{01}{01}{8494807} + $\Delta$\al\ and 
\dec{-1}{24}{31}{1293578} + $\Delta$\de. For B2252+172 the CRF coordinates
\Ra{22}{54}{59}{5974430} and \dec{17}{33}{24}{690713} (solution \#6, Table~\ref{t2icrf})
were used. The standard errors are the statistical standard errors
with a constant added in quadrature so that $\chinu = 1$.}
\label{t42252}
\end{deluxetable}

\subsection{Analysis of Geodetic Group-delay Observations of 3C 454.3 and B2250+194}

The data from two of our sources, 3C~454.3 and B2250+194, observed in
many of the geodetic VLBI sessions, each extending over about one day,
were analyzed by one of us (L. P.) with the VTD/post-Solve software package. 
We made a weighted
least-squares solution using all available geodetic VLBI observations
of 3955 sources, including \objectname[]{B2250+194}, \objectname[]{3C
454.3}, and the 212 ``defining" sources from the ICRF catalogue
\citep{FeyGJ2009}, made at 157 stations from 1979 to 2008 (dubbed
solution gpb$\_$2008a). All in all we used a total of 6.5 million
determinations of group delay from observations made simultaneously at
8.4 and 2.3 GHz. This solution forms our CRF. It is consistent with the 
ICRF2 \citep{FeyGJ2009}, which, however, does not provide information that we need as described below.

In particular, we estimated the coordinates of B2250+194 and 3C~454.3
for each observing session while forcing the coordinates of each other
source to be constant.  We also solved for the positions and
velocities of all stations, for polar motion and UT1 parameters, and
their rates of change, for nutation daily offsets at the middle epoch
of each session and for numerous other parameters such as those that
model clock offsets, atmosphere path delays in the zenith direction,
and tilts of the assumed atmosphere axis of symmetry.
We imposed the constraint that the net rotation of our estimates of
source positions of the 212 defining sources with respect to the
positions of these sources in the ICRF catalogue be zero.  Such a choice for the 
constraint provides the continuity of our solution to other VLBI 
solutions, including the ICRF2 solution. For more
details, see \citet{Petrov+2009}.

The individual source positions from this solution 
were estimated not with respect to a particular reference source, but
with respect to the entire ensemble of observed sources. This approach
was possible because the geodetic VLBI sessions were designed in such
a way that (1) a set of $\sim$100 core sources and $\sim$20 core
antenna sites (``stations'') were common to all sessions, and (2) the
resulting estimates of source positions, station positions, and Earth
orientation parameters would not be strongly correlated. The presence
of sources common to all sessions tended to ensure the consistency of
determination of the interferometer orientation with respect to these
core sources.

Although our list of sources has 3955 objects, the relative weight of
the $\sim$100 predominant sources ensured that they dominate the ensemble. Therefore, we interpret
the estimated positions of \objectname[]{B2250+194} and \objectname[]{3C 454.3} 
as positions almost entirely with respect to the ensemble of the $\sim$100 predominant extragalactic 
objects. 

For the solution yielding our CRF, each source was assumed to be a point
source. However, almost all of these sources exhibit some extended
structure at the milliarcsecond scale that may also vary over time.
Since these sources were not routinely imaged and hence the group
delays not corrected for structure effects, a position determination
from geodetic VLBI cannot be identified with respect to a specific
fiducial point in the brightness distribution of a source. That
fiducial point is therefore in principle unidentified for any source
with structure.  In general, however, the more compact the source, the
closer the estimated position is to the peak in the brightness
distribution of the source.
The source B2250+194 is sufficiently compact that for our purposes the
position determined is effectively that of the brightness peak.  For
3C~454.3 the fiducial point for each session is less well known
because of the complexity of the source structure and its changes with
time.  Moreover, the fiducial point could change due to changes of the
\uv\ coverage. For these reasons, we do not rely on the 3C~454.3
position from geodetic VLBI observations, but rather use our dedicated
VLBI observations and phase-referenced data to identify individual
components in the brightness distribution of 3C~454.3 and to determine
their positions relative to our two more compact extragalactic
sources, B2250+194 and B2252+172, which in turn we tie to the CRF.

\section{Astrometric Results (2): Fit for the Position at Epoch and Proper Motion}
\label{ast2}

\subsection{Fit: Components of 3C 454.3 Relative to B2250+194 and B2252+172}

\begin{figure}
\centering
\includegraphics[width=\textwidth]{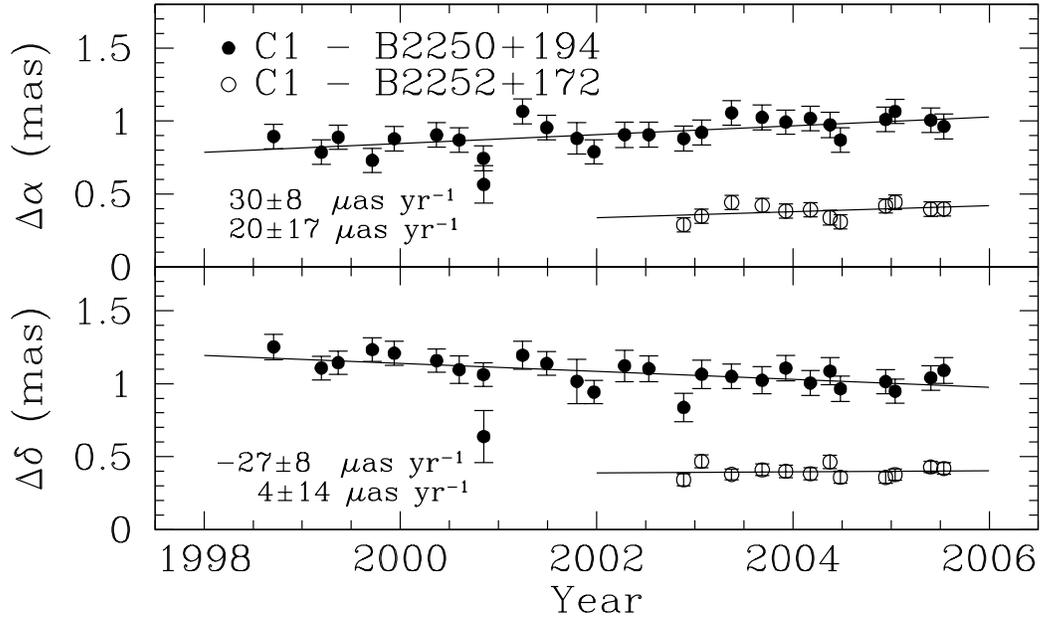}
\caption{The coordinates of C1 in 3C~454.3 relative to those of
B2250+194 and B2252+172 (except for offsets) as a function of
time. The values of $\Delta$\al\ and $\Delta$\de\ are obtained from
the entries in Tables~\ref{t32250} and \ref{t42252}. For discussion of
apparent partial correlations between the two position-difference data sets,
see text.}
\label{f9c15052}
\end{figure}

\begin{figure}
\centering
\includegraphics[width=\textwidth]{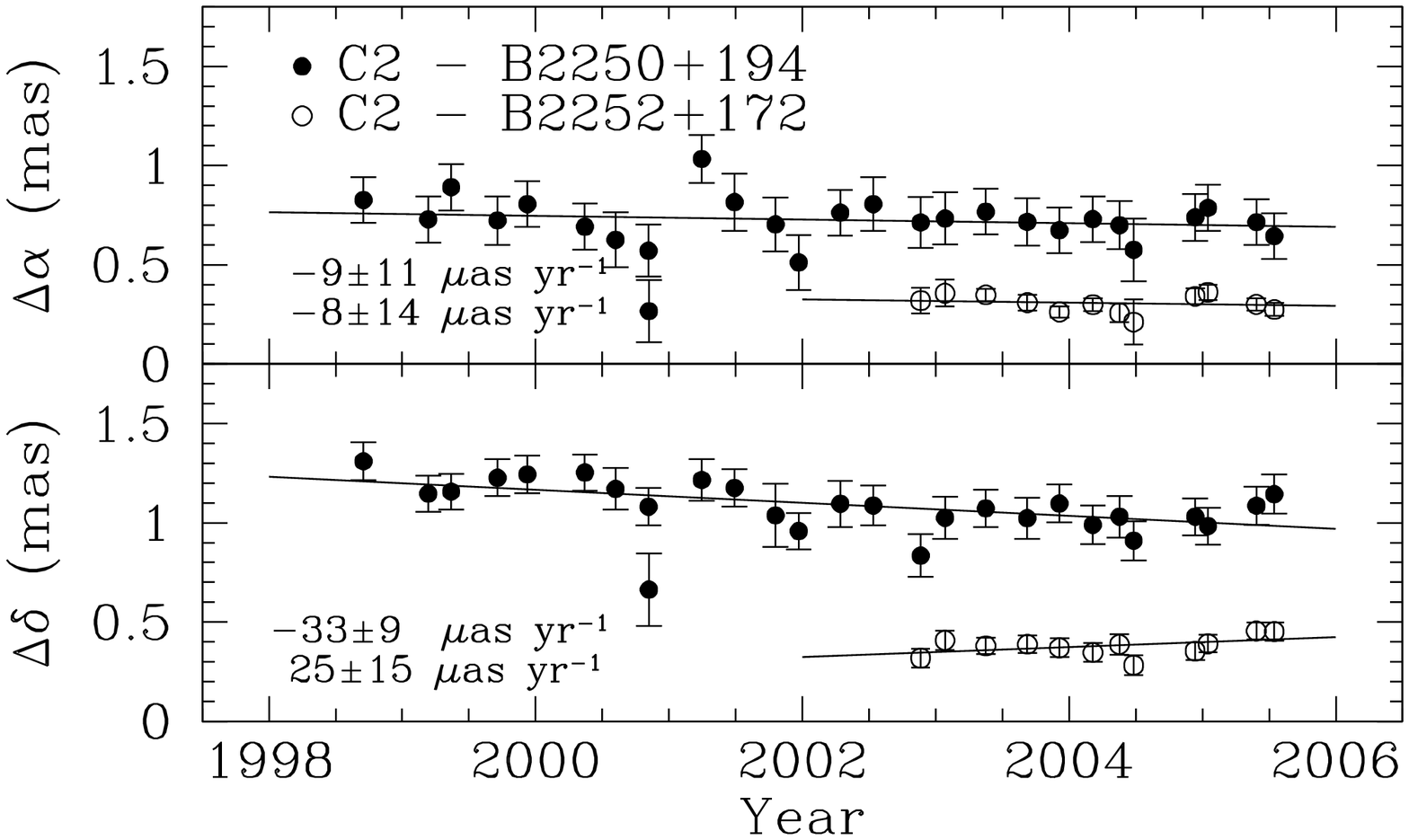}
\caption{As in Figure~\ref{f9c15052}, but now for C2 in 3C 454.3.}
\label{f10c25052}
\end{figure}

\begin{figure}
\centering
\includegraphics[width=\textwidth]{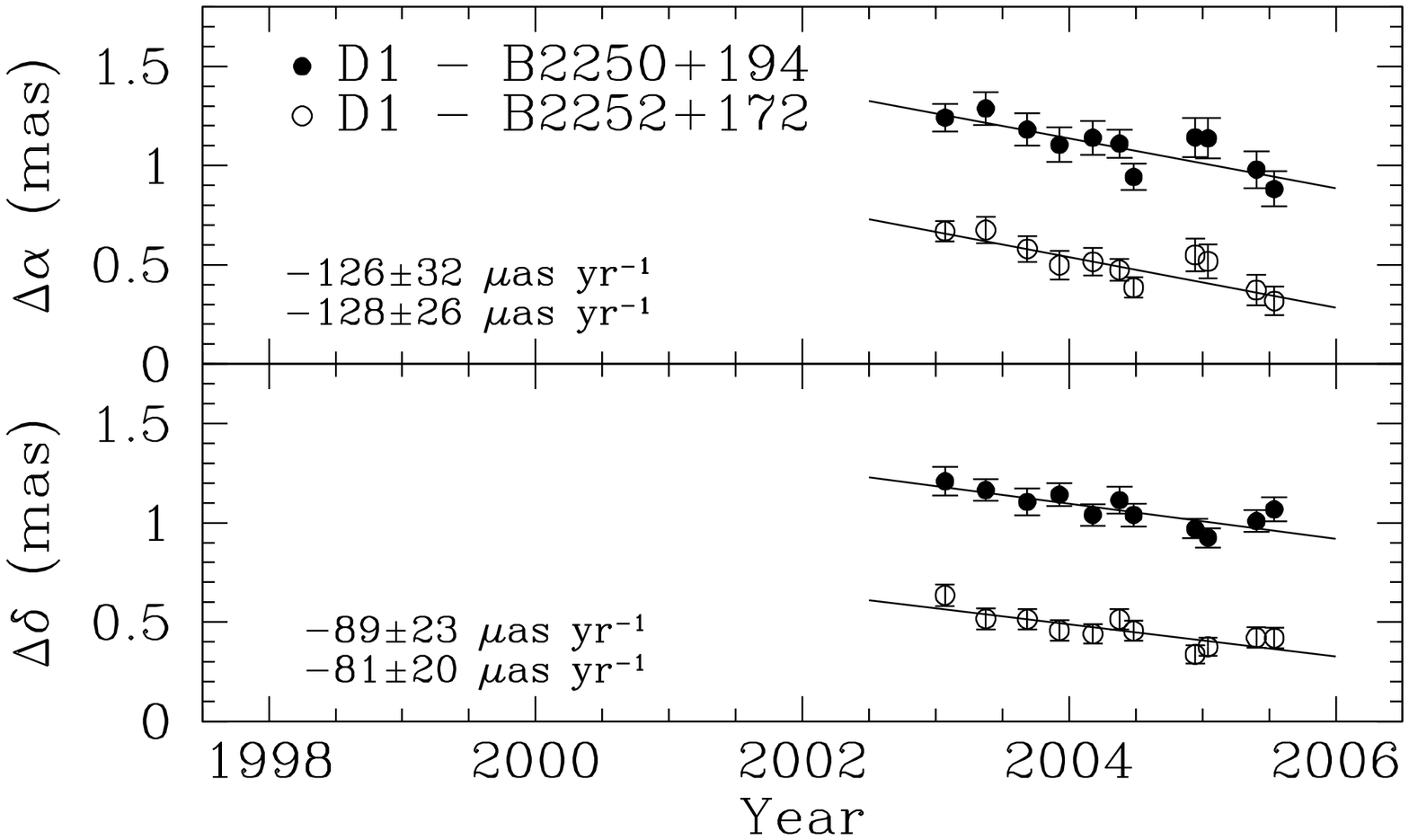}
\caption{As in Figure~\ref{f9c15052}, but now for D1 in 3C 454.3.}
\label{f11d15052}
\end{figure}

\begin{figure}
\centering
\includegraphics[width=\textwidth]{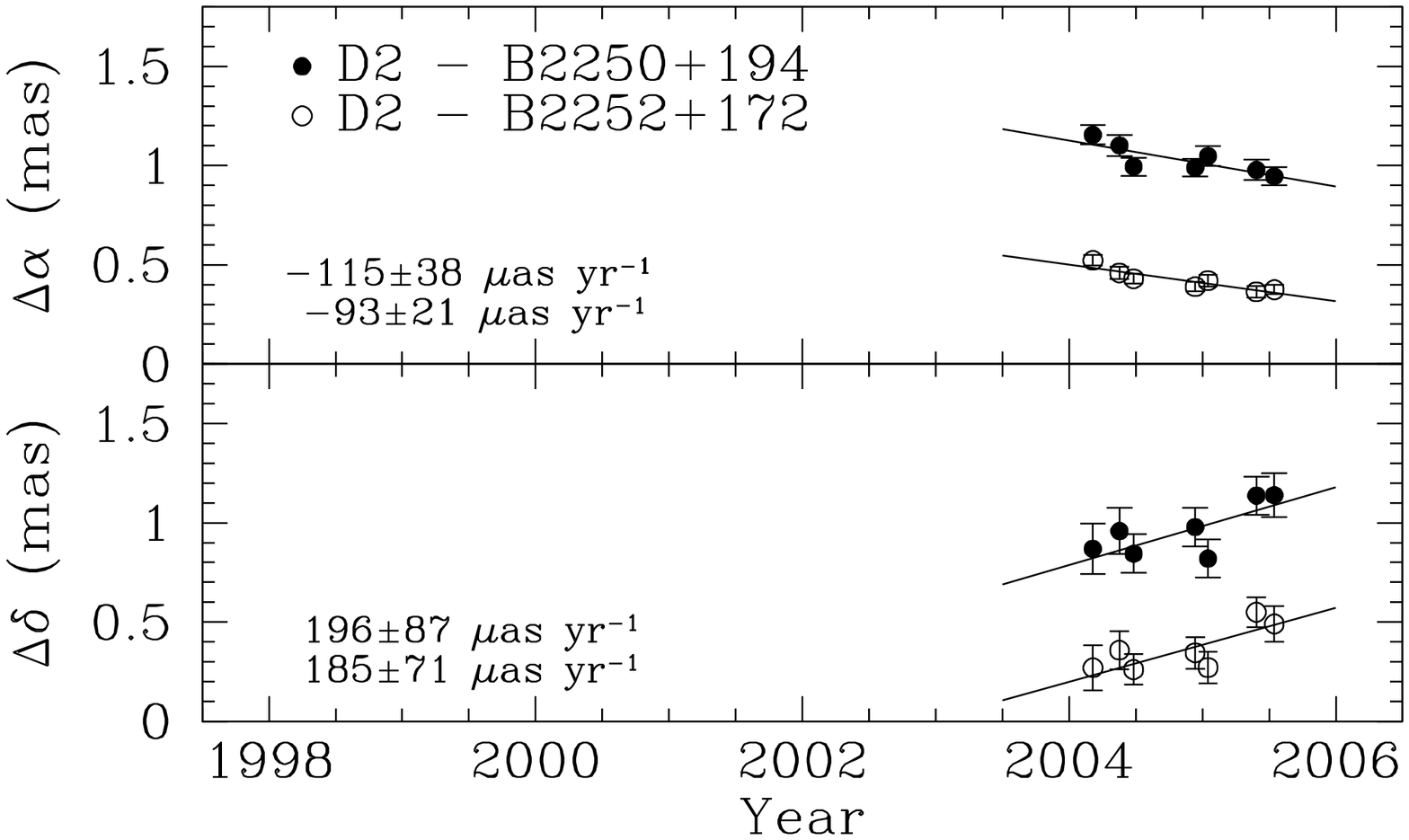}
\caption{As in Figure~\ref{f9c15052}, but now for D2 in 3C 454.3.}
\label{f12d25052}
\end{figure}

\begin{figure}
\centering
\includegraphics[width=\textwidth]{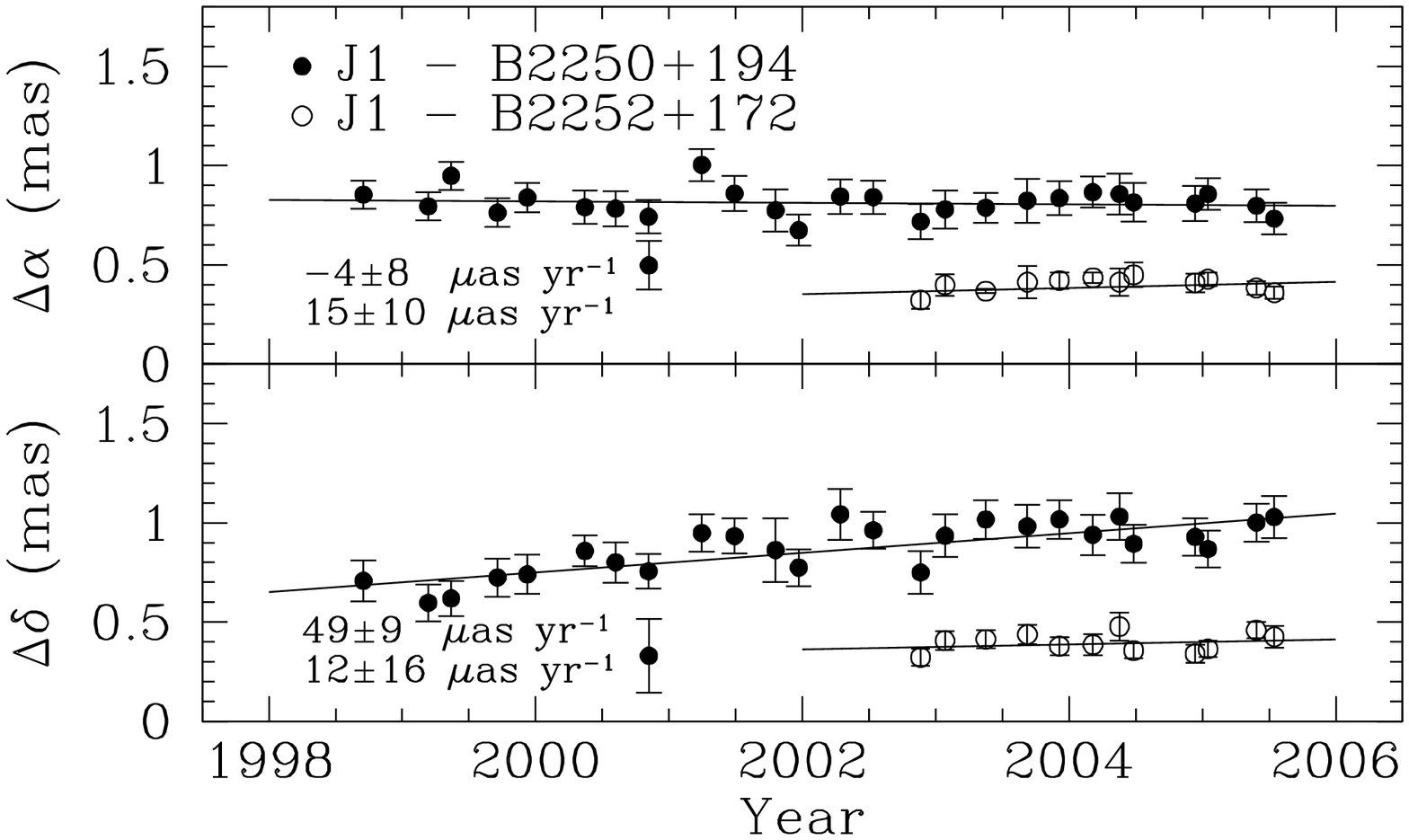}
\caption{As in Figure~\ref{f9c15052}, but now for J1 in 3C 454.3.}
\label{f13j15052}
\end{figure}

\begin{figure}
\centering
\includegraphics[width=\textwidth]{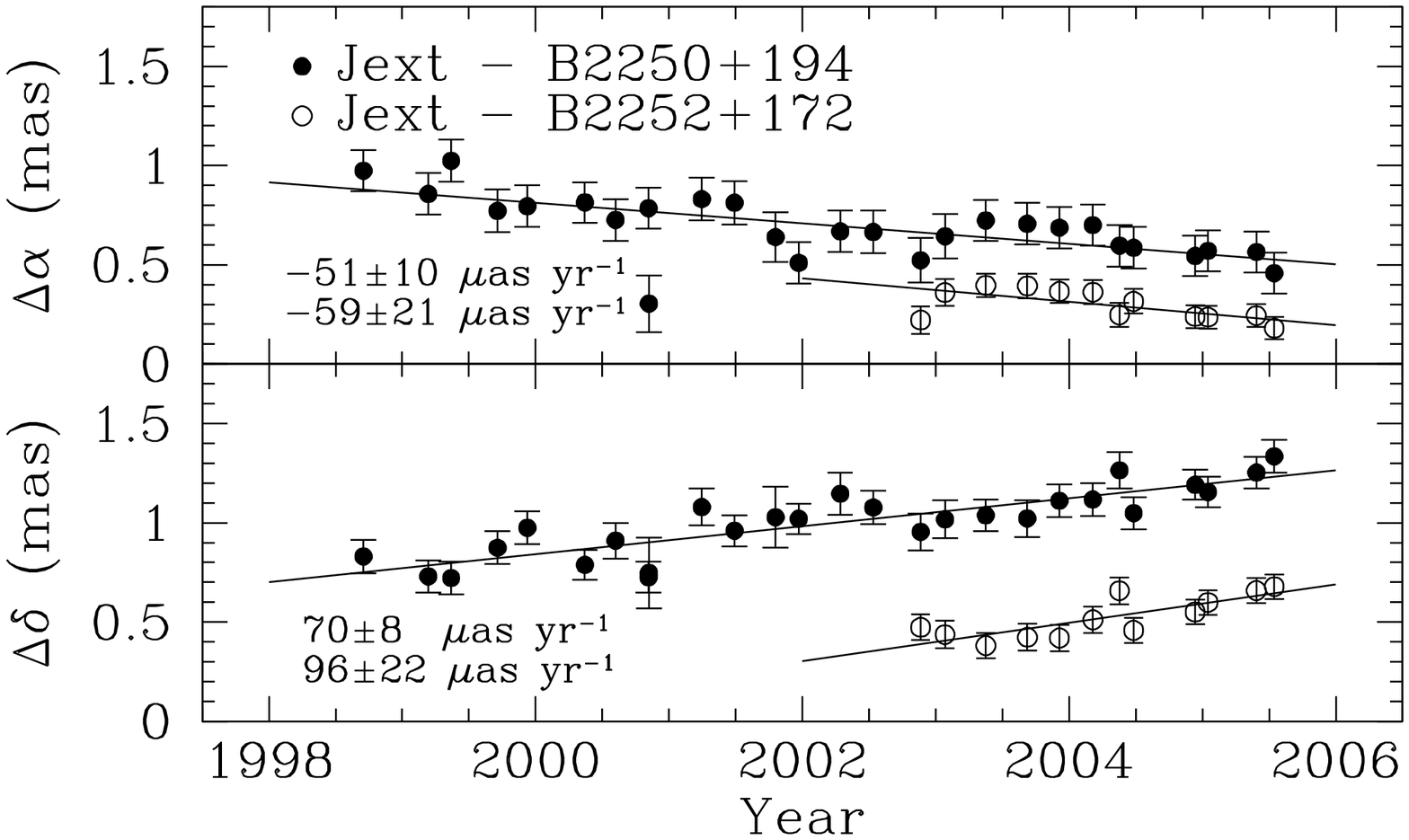}
\caption{As in Figure~\ref{f9c15052}, but now for Jext in 3C 454.3.}
\label{f14j25052}
\end{figure}

We determined the position at epoch and proper motion of each of C1,
C2, D1, D2, J1, and Jext relative to the brightness peaks of B2250+194
and B2252+172 with weighted least-squares fits for \al\ and separately
for \de. We list our results together with the weighted linear correlation
coefficients and the weighted post-fit rms values (wrms) in
Table~\ref{t5comp} (solutions \#1 and \#2).  We used the data
corrected with the JPL ionosphere model from Tables~\ref{t32250} and
\ref{t42252}.
The data and the corresponding lines from the fits (solutions \#1 and
\#2) are plotted in Figures~\ref{f9c15052}, \ref{f10c25052},
\ref{f11d15052}, \ref{f12d25052}, \ref{f13j15052}, and
\ref{f14j25052}. Not surprisingly, the smallest wrms values, in both
\al\ and \de, were obtained for the components of 3C~454.3 relative to
the close reference B2252+172 (solution \#2).
Our combined proper-motion estimates are given as solution
\#3. They were obtained as a weighted average of proper-motion
estimates relative to B2250+194 and B2252+172 for the same (short)
time range.

\subsection{Fit: B2252+172 Relative to B2250+194}

\begin{figure}
\centering
\includegraphics[width=\textwidth]{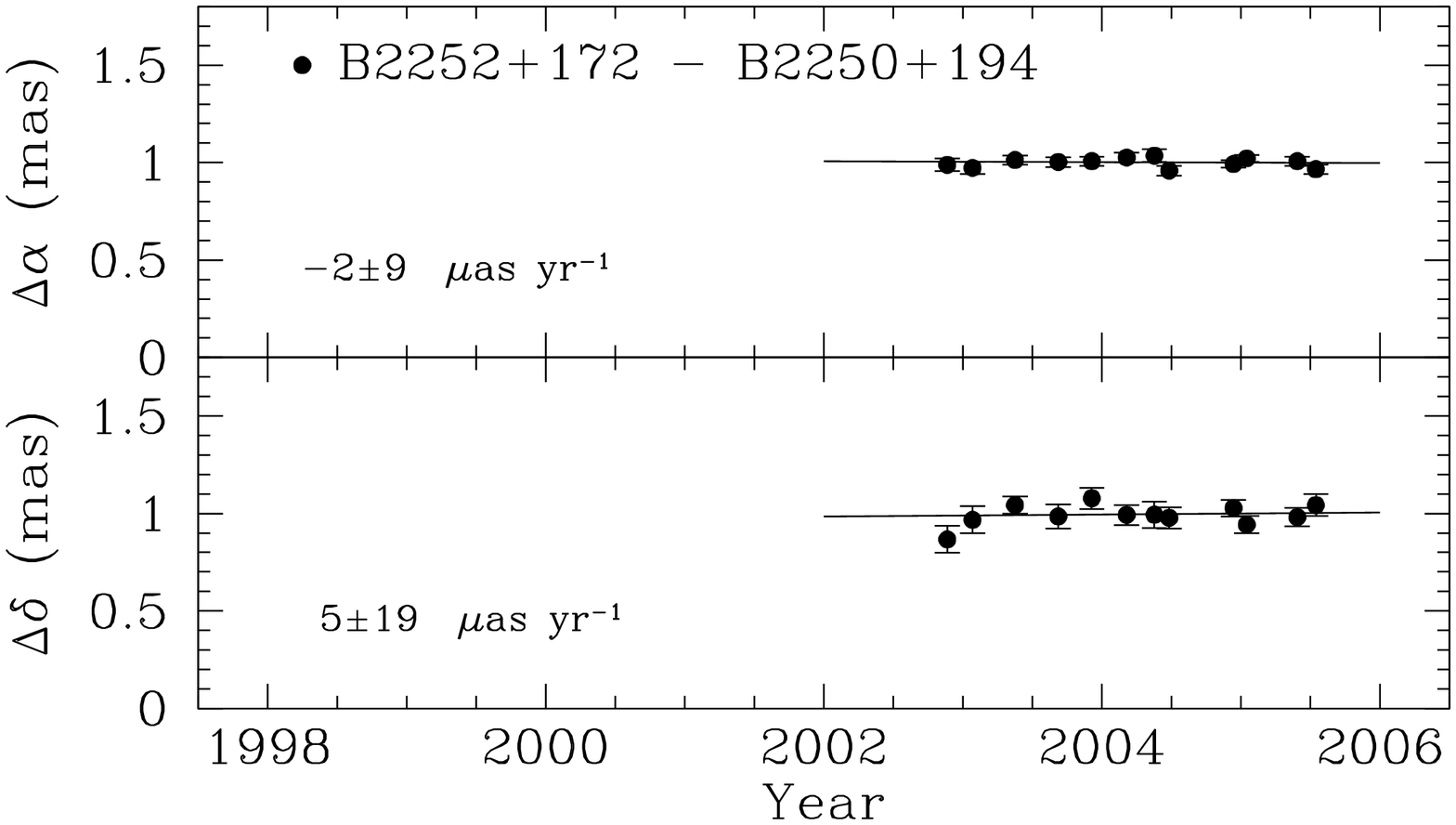}
\caption{The coordinates of B2252+172 relative to those of B2250+194
(except for an offset in each coordinate) as determined from the
differences of the position estimates, (2252-C1)$-$(2250-C1)
given in Tables~\ref{t32250} and \ref{t42252}.}
\label{f85250}
\end{figure}

The position at epoch and proper motion of B2252+172 relative to
B2250+194 were determined by first differencing the position
determinations of C1 relative to B2250+194 and C1 relative to
B2252+172 from Tables~\ref{t32250} and \ref{t42252} in the sense (C1
$-$ 2250) $-$ (C1 $-$ 2252)\footnote{Differencing the phase delays for
each scan at each epoch would have given us the position of B2252+172
relative to B2250+194 directly for each epoch and likely with a
somewhat smaller uncertainty. However, our procedure proved to also
give sufficiently accurate results for our purposes.}.  We then used
weighted least-squares to fit a straight line to these differences. We
list the results also in Table~\ref{t5comp} (solution \#4) and plot
the data with the fit line in Figure~\ref{f85250}.  The relative
proper motion of the two sources is zero within a small portion of 
1$\sigma$. The 1$\sigma$ upper limits are 11 and 24 \muasyr\ in \al\
and \de, respectively.

\begin{deluxetable}{c@{ }c@{ }c@{ }c r r r c c c r r@{}c@{}c@{}c}
\rotate
\tabletypesize{\tiny}
\tablecaption{Relative positions at epoch and proper motions of the components of 3C 454.3, B2250+194, and B2252+172\tablenotemark{a}}
\tablewidth{0pt}
\tablehead{
\colhead{Component} & \multicolumn{3}{c}{\al} & \colhead{$\mu_\alpha$}                & \colhead{$\rho_{\al}$} & \colhead{wrms$_{\al}$} & \multicolumn{3}{c}{\de} & \colhead{$\mu_{\delta}$} 
               & \colhead{$\rho_{\de}$}  & \colhead{wrms$_{\de}$} & \colhead{Range of epochs\tablenotemark{b}} & \colhead{Solution$\,$\#}  \\
\colhead{-reference}  & \colhead{(h)} &\colhead{(m)} &\colhead{(s)} &               \colhead{(\muasyr)}         &                        & \colhead{(\muas)}         & \colhead{(\arcdeg)} 
&\colhead{(\arcmin)} &\colhead{(\arcsec)} & \colhead{(\muasyr)} &                         & \colhead{(\muas)} 
}
\startdata
C1 $-$ 2250 & & &   50.3787837 (19)  &$30\pm\phn8$&   0.60  &  93  &$-$3& 33& 41.067505 (28) &$-27\pm\phn8$  &$-$0.56 &\phn85 & 1998.71 -- 2005.54 & 1\tablenotemark{c}  \\
C2 $-$ 2250 & & &   50.3787347 (27)  &$-9\pm11$   &$-$0.16  & 119  &$-$3& 33& 41.067591 (32) &$-33\pm\phn9$  &$-$0.58 &\phn97 & 1998.71 -- 2005.54 & 1\tablenotemark{c}  \\
D1 $-$ 2250 & & &   50.3786586 (25)  &$-126\pm32$ &$-$0.79  & 74   &$-$3& 33& 41.067913 (23) &$-89\pm23$     &$-$0.79 &\phn52 & 2003.07 -- 2005.54 & 1\tablenotemark{c}  \\
D2 $-$ 2250 & & &   50.3784787 (14)  &$-115\pm38$ &$-$0.80  & 41   &$-$3& 33& 41.067960 (42) & $196\pm87$      & 0.71 &\phn88   & 2004.18 -- 2005.54 & 1\tablenotemark{c}  \\
J1 $-$ 2250 & & &   50.3783932 (19)  &$-4\pm\phn8$&$-$0.10  &  80  &$-$3& 33& 41.066488 (33) & $49\pm\phn9$    & 0.73 &\phn96 & 1998.71 -- 2005.54 & 1\tablenotemark{c}  \\
Jext $-$ 2250 &&&   50.3783141 (24)  &$-51\pm10$  &$-$0.72  & 103  &$-$3& 33& 41.065042 (27) & $70\pm\phn8$    & 0.87 &\phn83   & 1998.71 -- 2005.54 & 1\tablenotemark{c}  \\
\\
C1 $-$ 2252 & &$-$01& 01.8494857 (14)&$20\pm17$   & 0.36    &  45  &$-$1& 24& 31.129358 (16) & $4\pm14$       & 0.09 &\phn37   & 2002.89 -- 2005.54 & 2\tablenotemark{d}  \\
C2 $-$ 2252 & &$-$01& 01.8495343 (10)&$ -8\pm14$  &$-$0.19  &  35  &$-$1& 24& 31.129424 (18) & $25\pm15$       & 0.46 &\phn42   & 2002.89 -- 2005.54 & 2\tablenotemark{d}  \\
D1 $-$ 2252 & &$-$01& 01.8496108 (20)&$-128\pm26$ &$-$0.86  &  59  &$-$1& 24& 31.129789 (20) &$-81\pm20$     &$-0.80$ &\phn46 & 2003.07 -- 2005.54 & 2\tablenotemark{d}  \\
D2 $-$ 2252 & &$-$01& 01.8497412 (08)&$-93\pm21$  &$-$0.89  &  23  &$-$1& 24& 31.129750 (35) &$186\pm72$      & 0.76 &\phn72   & 2004.18 -- 2005.54 & 2\tablenotemark{d}  \\
J1 $-$ 2252 & &$-$01& 01.8498757 (19)&$ 15\pm10$  & 0.43    &  26  &$-$1& 24& 31.128410 (19) &$12\pm16$       & 0.24 &\phn43   & 2002.89 -- 2005.54 & 2\tablenotemark{d}  \\
Jext $-$ 2252&&$-$01& 01.8499544 (16)&$-59\pm21$  &$-$0.67  &  55  &$-$1& 24& 31.126911 (25) &$96\pm22$       & 0.81 &\phn59   & 2002.89 -- 2005.54 & 2\tablenotemark{d}  \\
\\
C1 $-$ 2250/2252 &&&                 & $20\pm13$  &         &      &   &   &                &  $7\pm12$       &      &       & 2002.89 -- 2005.54 & 3\tablenotemark{e}  \\
C2 $-$ 2250/2252 &&&                 &$-10\pm11$  &         &      &   &   &                & $28\pm14$       &      &       & 2002.89 -- 2005.54 & 3\tablenotemark{e}  \\
D1 $-$ 2250/2252 &&&                 &$127\pm20$  &         &      &   &   &                &$-84\pm15$       &      &       & 2002.89 -- 2005.54 & 3\tablenotemark{e}  \\
D2 $-$ 2250/2252 &&&                 &$-98\pm18$  &         &      &   &   &                &$190\pm56$       &      &       & 2002.89 -- 2005.54 & 3\tablenotemark{e}  \\
J1 $-$ 2250/2252 &&&                 &$12\pm\phn8$&         &      &   &   &                & $15\pm14$       &      &       & 2002.89 -- 2005.54 & 3\tablenotemark{e}  \\
Jext $-$ 2250/2252 &&&               &$-59\pm16$  &         &      &   &   &                &$104\pm15$       &      &       & 2002.89 -- 2005.54 & 3\tablenotemark{e}  \\
\\
\\
2252 $-$ 2250&00 & 01 & 52.2282694 (6) &\phn-$2\pm 9$ &$-$0.06    &  22 &  $-$2& 09 & 09.938073 (20) & $5\pm19$        & 0.08 &\phn48    & 2002.89 -- 2005.54   & 4\tablenotemark{f} \\
\enddata
\tablenotetext{a}{The coordinates and proper motions of the core and
jet components of 3C~454.3 for the reference epoch -- the midpoint of the \GPB\ mission -- 2005 Feb.\ 1
(2005.08), using VLBI differential observations of 3C~454.3 relative
to B2250+194 and B2252+172.  The parameters are derived from weighted
least-squares linear fits.  The uncertainties are standard errors
derived from the fit on the basis of the statistical standard errors
of the individual measurements added in quadrature to constants so
that $\chinu = 1$. We searched for any sign that a possible
correlation between consecutive data points (see, e.g.,
Fig~\ref{f9c15052}) could render our uncertainties too small by
repeating the weighted least squares fit for even and odd numbered
data points separately.  However, we did not find any such sign and
therefore think that any possible correlation would only have a minor
effect on our uncertainty estimates.  Also listed are the weighted linear correlation
coefficients, $\rho$, from the fits, and the weighted
rms values, wrms, for the post-fit residuals, for the two
coordinates. All data are corrected for the effects of the ionosphere
by using a model implemented in the AIPS software and based on GPS
data (JPL, see text). The earliest date at which the model could be
used is 1998 Sept. 17 (1998.71).  For D1 and D2 the time range starts
with the first epoch at which a component could be identified in the
brightness distribution, that is, 2003 Jan. 26 (2003.07) and 2004 Mar.
6 (2004.18), respectively.}
\tablenotetext{b}{The range of epochs from which the data were taken
for the solution. The range 1998.71 -- 2005.54 is the time range from
1998 Sep.\ 17 to 2005 July 16 of our \GPB\ VLBI observations for which
the ionospheric model, JPL, could be used. The epoch 2002.89 refers to
2002 Nov.\ 20 when we started to include B2252+172 in our
observations.  The other time ranges refer to the total ranges for
which data were collected for the respective component and source. The reference
epoch 2005.09 is 2005 Feb.\ 1, the midpoint of the time period through
which data were taken on the \GPB\ spacecraft.}
\tablenotetext{c}{The differences from B2250+194 in the coordinates of
the positions and in the corresponding components of the proper motion
for each of 3C~454.3's six core and jet components. The time range
starts for all components with the first epoch of our VLBI
observations, except for components D1 and D2, where it starts with the
first epoch at which a component could be identified in the brightness
distribution.  The epochs of the time range are 1998 Sept. 17
(1998.71), 2003 Jan. 26 (2003.07), and 2005 Jul. 16 (2005.54).}
\tablenotetext{d}{As in c for solution \#1 but now relative to
B2252+172. The time range starts with the date of the first B2252+172
observations, 2002 Nov.\ 20 (2002.89).}
\tablenotetext{e}{As for solution \#1 but now starting not earlier
than at epoch 2002.89, and including the values for B2252+172.  In
particular, for each component, we took the weighted average of (a)
the proper motion from a solution (3C~454.3-2250) for the short time
range (not listed), and (b) the proper motion of solution \#2
(3C~454.3-2252).
The errors from (a) and (b) were added in quadrature.}
\tablenotetext{f}{The coordinates of B2252+172 relative to those of
B2250+194 and their changes with time. They were derived from the data
(2252 $-$ C1) $-$ (2250 $-$ C1) determined for each epoch.}
\label{t5comp}
\end{deluxetable}

\subsection{Fit: 3C 454.3, B2250+194, and B2252+172 in the CRF}

\subsubsection{3C 454.3}
\begin{figure}
\centering
\includegraphics[width=\textwidth]{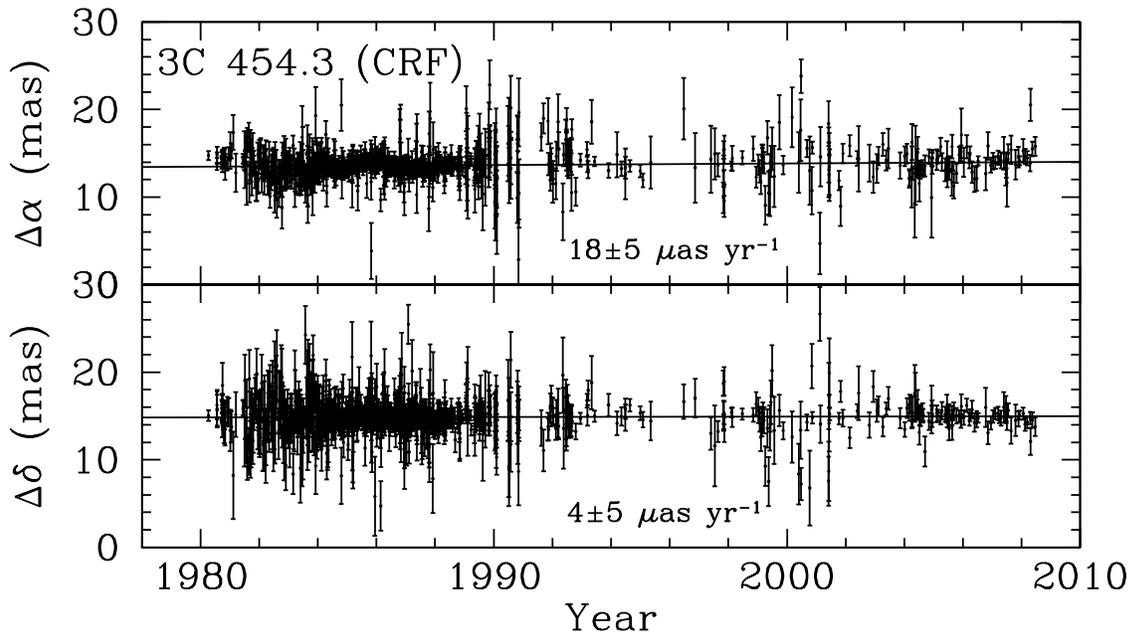}
\caption{The coordinates of 3C~454.3 as determined from
routine geodetic VLBI observations of up to $\sim$4000 extragalactic sources
scattered over the sky. 
Here
and hereafter for plots of coordinates versus time, the straight lines
give the weighted least-squares fit to the data points.  The slopes of
the lines and their statistical standard errors are indicated. These
errors are standard errors adjusted with a constant added in
quadrature to the statistical standard errors for each coordinate so
that $\chinu=1$ (see text).}
\label{f53cicrf}
\end{figure}

In Figure~\ref{f53cicrf} we plot the position determinations of
3C~454.3 from geodetic VLBI observations spanning almost 30 years.
The statistical standard errors of the coordinates vary widely in part
because of the different lengths of time this source was observed in
the various sessions, and in part because the sensitivity of the VLBI
systems used for the observations generally improved over time.

We solved for the position and proper motion of 3C~454.3  
by using weighted least-squares fits for \al\ and \de\ separately, 
for the whole time range of observations from 1980 to 2008.
Since these weighted least-squares fits to the data gave $\chinu$ larger than unity,
we again added a constant in quadrature to the statistical standard
errors, separately for \al\ and \de, so as to obtain $\chinu=1$ for
each coordinate.  The standard errors for the individual data points
so determined, together with the fit lines, are plotted in
Figure~\ref{f53cicrf} and the results are listed in
Table~\ref{t2icrf} (solution \#1). 
In our CRF, the proper-motion component in \de\ is zero within
0.8$\sigma$. However, in \al\ it is $18\pm5$~\muasyr, non-zero at a
3.6$\sigma$ significance level.
However, since this determination does not refer consistently to any particular
component in the brightness distribution of the source, its significance
is not clear.

\subsubsection{B2250+194}
\begin{figure}
\centering
\includegraphics[width=\textwidth]{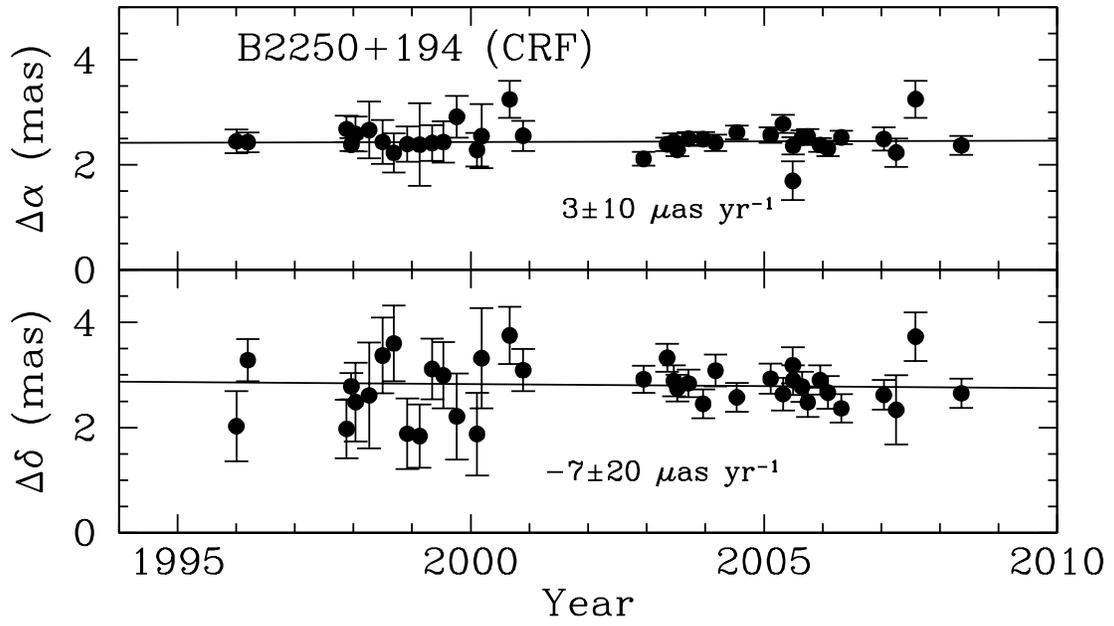}
\caption{The change in coordinates of B2250+194\@.  Otherwise similar
to Figure~\ref{f53cicrf}.}
\label{f650icrf}
\end{figure}

The second of our sources observed with geodetic VLBI is B2250+194.
About 12 years of such observations yielded determinations of position
and proper motion of B2250+194 in the CRF. Via this tie and our phase-delay observations, 
we determined the positions of the components of 3C~454.3, of B2252+172, and of IM
Peg in the CRF.  For the source B2250+194, the standard errors of the position
estimates also vary widely from session to session because of the
different spans of time over which observations of this source were
spread in the various sessions. As for 3C~454.3, we plot the
coordinates for B2250+194 from the different observing sessions and
their standard errors, computed as described above, together with the
linear fit lines (solution \#2 in Table~\ref{t2icrf}) 
for the whole observing period, from 1997 to 2008, in
Figure~\ref{f650icrf}.  
The proper motion is zero within 0.4$\sigma$.

\subsubsection{B2252+172}

Using the position at epoch and proper motion of B2250+194 in the CRF,
given in Table~\ref{t2icrf} (solution \#2), we also obtain the
equivalent values for B2252+172 in the CRF\@. In particular, we take
the values from the above solution \#2 and add them to the values from
our phase-reference observations from Table~\ref{t5comp} (solution
\#4) in the sense 2250 + (2252 $-$ 2250) and list them also in
Table~\ref{t2icrf} (solution \#3). The proper motion is zero within
0.1$\sigma$.

Owing to a recent VLBA sensitivity upgrade, it became possible now 
to determine group delays of sources as weak as 10~mJy. Source B2252+172 
was observed in two scans each 320~s long, with a 9-station VLBA network 
at epoch  2011 August 14 at 8.4 GHz during a gap in the schedule of target sources in 
a VLBI survey of bright 
infrared galaxies \citep{Condon+2011}. 
The source appeared slightly resolved with a correlated 
flux density in a range of 11--13 mJy at a range of baseline projection 
lengths 3--200 megawavelengths. Its group-delay coordinates from that 
experiment are $22^{\rm h} 54^{\rm m} 59\fs{597\,449}  \pm 0.000\,015, \enskip 
+17\arcdeg 33\arcmin 24\farcs{690\,49} \pm 0.000\,42$. These coordinates
agree with those from Table~\ref{t5comp} (solution \#3) within the formers sixfold larger uncertainties. The 
corresponding proper motion for the 6.6 years between our reference epoch (see Table~\ref{t5comp}) 
and the observing date in 2011 is $13\pm33~\muas$ and $-34\pm64~\muas$, which gives $1\sigma$ upper
limits about threefold larger than those in Table~\ref{t5comp}.

\begin{deluxetable}{c c c c r c c c c c r r c c c}
\rotate
\tabletypesize{\tiny}
\tablecaption{The positions at epoch and proper motions in the celestial reference frame (CRF)\tablenotemark{a}}
\tablewidth{0pt}
\tablehead{
\colhead{Component-reference} & \multicolumn{3}{c}{\al} & \colhead{$\mu_\alpha$}                & \colhead{$\rho_{\al}$} & \colhead{wrms$_{\al}$} & \multicolumn{3}{c}{\de} & \colhead{$\mu_{\delta}$} 
               & \colhead{$\rho_{\de}$}  & \colhead{wrms$_{\de}$} & \colhead{Range of epochs\tablenotemark{b}} & \colhead{Solution \#}  \\
 & \colhead{(h)} &\colhead{(m)} &\colhead{(s)} &               \colhead{(\muasyr)}         &                        & \colhead{(\muas)}         & \colhead{(\arcdeg)} 
&\colhead{(\arcmin)} &\colhead{(\arcsec)} & \colhead{(\muas~yr)} &                         & \colhead{(\muas)}    &                             & 
}
\startdata
3C 454.3  & 22& 53 & 57.7479664 (60)  & $18\pm\phn5$&   0.11 & 742  & 16 & 08 & 53.560943 (80)&$\phn4\pm\phn5$& 0.03  &\phn918  & 1980.28 -- 2008.47  & 1  \\
B2250+194 & 22& 53 &\phn7.3691736 (24)&    $3\pm10$ & 0.05   & 179  & 19 & 42 & 34.628786 (62) &$-7\pm20$     &$-0.06$&\phn345  & 1996.01 -- 2008.36  & 2  \\ 
B2252+172 & 22& 54 & 59.5974430 (25)  &    $1\pm13$ &        &      & 17 & 33 & 24.690713 (65) &$-2\pm28$     &       &         & 2002.89 -- 2005.54  & 3\tablenotemark{c}  \\ 
\\
C1 (50)  &22&53& 57.7479573 (31)      &$33\pm13$    &\nodata &\nodata & 16 & 08 & 53.561281 (68) &$-35\pm21$    & \nodata  & \nodata  & 1998.71 -- 2005.54 & 4\tablenotemark{d}  \\
C2 (50)  &22&53& 57.7479083 (36)      &$-7\pm15$    &\nodata &\nodata & 16 & 08 & 53.561195 (70) &$-40\pm22$    & \nodata  & \nodata  & 1998.71 -- 2005.54 & 4\tablenotemark{d}  \\
D1 (50)  &22&53& 57.7478322 (35)      &$-123\pm34$  &\nodata &\nodata & 16 & 08 & 53.560873 (66) &$-96\pm30$    & \nodata  & \nodata  & 2003.07 -- 2005.54 & 4\tablenotemark{d}  \\
D2 (50)  &22&53& 57.7476523 (28)      &$-112\pm39$  &\nodata &\nodata & 16 & 08 & 53.560826 (75) &$188\pm89$    & \nodata  & \nodata  & 2004.18 -- 2005.54 & 4\tablenotemark{d}  \\
J1 (50)  &22&53& 57.7475668 (31)      &$-1\pm13$    &\nodata &\nodata & 16 & 08 & 53.562298 (70) & $42\pm22$    & \nodata  & \nodata  & 1998.71 -- 2005.54 & 4\tablenotemark{d}  \\
Jext (50) &22&53& 57.7474877 (34)     &$-49\pm14$   &\nodata &\nodata & 16 & 08 & 53.563744 (68) & $63\pm21$    & \nodata  & \nodata  & 1998.71 -- 2005.54 & 4\tablenotemark{d}  \\
\\
C1 (50/52)     &\nodata&\nodata&\nodata&$22\pm17$    &\nodata&\nodata&\nodata&\nodata& \nodata&  $4\pm26$    &\nodata&\nodata& 2002.89 -- 2005.54 & 5\tablenotemark{e}  \\
C2 (50/52)     &\nodata&\nodata&\nodata&$-8\pm15$    &\nodata&\nodata&\nodata&\nodata& \nodata& $25\pm27$    &\nodata&\nodata& 2002.89 -- 2005.54 & 5\tablenotemark{e}  \\
D1 (50/52)     &\nodata&\nodata&\nodata&$-125\pm23$  &\nodata&\nodata&\nodata&\nodata& \nodata&$-92\pm27$    &\nodata&\nodata& 2003.07 -- 2005.54 & 5\tablenotemark{e}  \\
D2 (50/52)     &\nodata&\nodata&\nodata&$-99\pm25$   &\nodata&\nodata&\nodata&\nodata& \nodata&$185\pm59$    &\nodata&\nodata& 2004.18 -- 2005.54 & 5\tablenotemark{e}  \\
J1 (50/52)     &\nodata&\nodata&\nodata& $13\pm14$   &\nodata&\nodata&\nodata&\nodata& \nodata& $10\pm24$    &\nodata&\nodata& 2002.89 -- 2005.54 & 5\tablenotemark{e}  \\
Jext (50/52)   &\nodata&\nodata&\nodata&$-58\pm25$   &\nodata&\nodata&\nodata&\nodata& \nodata&$100\pm26$    &\nodata&\nodata& 2002.89 -- 2005.54 & 5\tablenotemark{e}  \\

\enddata
\tablenotetext{a}{The source coordinates, $\alpha$ and $\delta$, of 3C
  454.3, B2250+194, and B2252+172 at the reference epoch 2005.08, and
  their changes with time, $\mu_\alpha$ and $\mu_\delta$ respectively,
  in the CRF.  The parameters are derived from weighted least-squares
  linear fits.  The uncertainties are standard errors derived from the
  fit on the basis of the statistical standard errors of the
  individual measurements added in quadrature to constants so that
  $\chinu = 1.$.  Also listed are the weighted linear correlation coefficients, $\rho$, 
from the fits, as well as the weighted rms
  values, wrms, for the post fit residuals, for the two coordinates.}

\tablenotetext{b}{The range of epochs as in Table~\ref{t2icrf}.}

\tablenotetext{c} {The coordinates of B2252+172 and their changes with
time derived by adding the solutions from \#2 to those of \#4 in Table~\ref{t5comp}. The
errors were added in quadrature.}

\tablenotetext{d}{The coordinates and proper motions of the components of 3C 454.3 in the CRF.
We added the values  from
solution \#1 of Table~\ref{t5comp} to the values from solution \#2 of this table in the sense 
(3C~454.3 comp. $-$ 2250) +2250, and added the errors in quadrature.}

\tablenotetext{e}{As for solutions \#4 but now only for the proper
motion, and at epochs not earlier than 2002.89, and with the inclusion
of the values for B2252+172.  In particular, we added the values from
solution \#3 of Table~\ref{t5comp} to the values from solution \#2 of
this table in the sense (3C~454.3 comp. $-$ 2250/2252) + 2250 and
added the errors in quadrature.}

\label{t2icrf}
\end{deluxetable}

\subsection{Fit: Components of 3C 454.3 in the CRF}

We determined the position at epoch and proper motion of each of C1,
C2, D1, D2, J1, and Jext in the CRF by combining the values relative
to B2250+194 and B2252+172 with the values of the latter sources in
the CRF.  In particular for the positions, we take the values from
Table~\ref{t5comp} (solution \#1) and add them to the values from
Table~\ref{t2icrf} (solution \#2) in the sense (3C~454.3 components
$-$ 2250) + 2250.  For the proper motion values we include the
data involving B2252+172 since they are independent of those involving
B2250+194. We thus take all data into account, but do so only for
the restricted range of epochs for which we have B2252+172 VLBI
observations.  In particular, we took the average solution \#3 from
Table~\ref{t5comp} and added to it the values of B2250+194 in the CRF
by using the solution \#2 from Table~\ref{t2icrf} in
the sense (3C~454.3 components $-$ 2250/2252) + 2250.  We list the resulting
position-at-epoch and proper-motion values for each component in
Table~\ref{t2icrf} (solution \#4).

\subsubsection{Position of the Core Component, C1, of 3C 454.3}

We now discuss position determinations obtained
via our two different astrometric
techniques: the position of 3C~454.3 in the CRF obtained from geodetic
group-delay observations (solution \#1, Table~\ref{t2icrf}) and the
position of C1 in the CRF obtained through a combination of geodetic
group-delay and our phase-delay observations (solution \#4,
Table~\ref{t2icrf}).
We emphasize that these two estimates are not expected to coincide, even 
in principle, because the former relates to some (ill defined) average 
over source structure and the latter to a far better defined component (C1) 
within that structure.
The difference between the two position determinations (pure group
delay minus combination) is $9.1 \pm6.8$~\mus\
in \al\ and $-338 \pm 105$~\muas\ in \de. While the difference in \al\
is only 1.3$\sigma$, the difference in \de\
is 3.2$\sigma$, large enough to perhaps be significant. 

In this
context we compare our pure group-delay position
estimate for 3C~454.3 with other such estimates. For instance,
the ICRF2 catalog \citep{FeyGJ2009} provides position estimates of 3C 454.3 
and B2250+194\footnote{This source is incorrectly referred to as B2250+190 in the 
ICRF2 catalog.} which are close to ours. It is based on essentially the same set of 
group delays, the same data editing, and the same software as we used but with a slightly 
different reduction and estimation model.
The differences (ours $-$ ICRF2) are:
$-16\pm62$ \mus\ and $-0.2\pm3.5$ \mus\ in \al\ and $-4\pm970$ \muas\ and $39\pm62$ \muas\ in \de, for 3C~454.3 
and B2250+194, respectively. Here we take our estimate of an apparent proper motion in \al\ of $18\pm5$ \muasyr\
into account and propagate our estimate back 11.13 years to their mean epoch of 1993.95. Our estimate 
of the proper motion in \de\ and the corresponding estimates for B2250+194 were small enough so that it 
was not necessary to consider them. The uncertainties are those from the ICRF2 listings only.

Further, the latest estimate from the USNO celestial reference frame solution,
crf2009b\footnote{http://rorf.usno.navy.mil/vlbi/}, which uses essentially the same data
set as we used only extended by another year, is
different from ours in the sense (ours $-$ USNO) by $-1.7\pm6.1$~\mus\ and $-1.4\pm1.9$~\mus\
in \al\ and $-21\pm17$ \muas\  and $52\pm31$ \muas\ in \de, for 3C~454.3 and B2250+194, respectively. 
Here again we take our estimate of an apparent proper motion in \al\ for 3C~454.3 into account and 
propagate our estimate back 17.79 years to 
their mean epoch of 1987.29. The uncertainty is our proper motion error added in quadrature with 
the USNO position error.
For the other differences, the proper motion estimates did not need to be taken into account. 
The uncertainties are those
from the USNO position estimates only.

Another
estimate of the position of 3C~454.3, but not of B2250+194, was made
recently with group-delay observations at 24 GHz
\citep{Lanyi+2010}. 
The corresponding differences are $-15.6\pm8$
\mus\ and $-177\pm176$ \muas. With our, statistically independent,
errors added in quadrature, the difference in \al\ reduces to
$1.6\sigma$ and in \de\ to $0.9\sigma$. We do not consider these
differences to be significant. Nevertheless, note that the difference in 
\de\ is in the direction to reduce the difference with our determination 
for C1.  

Since we compare here results from different catalogs, we point out that they are almost 
identical in their overall orientations. The differences in these orientations 
correspond to a level of only several tens of microarcseconds, which is negligible for our purposes.

To summarize:
First there is a $-338$ \muas\ ($3.2\sigma$) difference in \de\ between our pure 
group-delay position of 3C 454.3 and the combined
group-delay phase-delay position of C1. Second, our pure group-delay position 
determinations for 3C 454.3 and, for comparison, also for B2250+194 agree within 
$<1.7\sigma$ with the USNO and ICRF2 position determinations.
Third, the position of 3C~454.3 at 24 GHz in \de, while its error is
large, cuts the above 3C~454.3/C1 discrepancy in half.  We discuss
these results below in \S~\ref{discuss}.

\subsubsection{Limit on the Proper Motion of the Core Component C1 of 3C 454.3}
\label{limit}

Table~\ref{t5comp} (solution \#1) shows that from 1998 to 2005 the
proper motion of C1 relative to B2250+194 is in the southeast
direction at a significance level of 3.8$\sigma$ and 3.4$\sigma$ in \al\ and
\de, respectively. In contrast, from 2002 to 2005 the motion of C1
relative to either B2250+194 (solution not listed) or B2252+172
(solution \#2) is smaller, and with an uncertainty corresponding to $<1.2~\sigma$, not
significant, although the errors are larger.  In any case, it appears
that either \objectname[]{B2250+194} moved to the northwest or C1
indeed moved to the southeast, particularly during the time from 1998
to 2002.

The cause of this apparent motion is not clear.  Is it possible that
the brightness peak of B2250+194 belongs to a jet that moved
northwestward during the early time period?  Figure~\ref{f32250} shows
that \objectname[]{B2250+194} is elongated to the northwest and also
slightly to the south, with the brightness peak located near the
center of its curved structure. If there were a supermassive black
hole located at the southern end of the structure, then the brightness
peak would likely belong to a jet component moving away from the black
hole, in this case to the northwest.  The geodetic observations which
determined the proper motion of B2250+194 in the CRF show an
insignificant northward motion of $21\pm46$ \muasyr\ for the time from
1998 to 2005 (solution not listed).  The observations over the longer
period from 1996 to 2008 are collectively more sensitive, but do not
indicate any motion to the northwest.  Moreover, the geodetic
observations of 3C~454.3 itself are too insensitive for a useful
proper-motion determination for the period from 1998 to 2005.

A useful way to test this jet hypothesis would be to determine the
position of the brightness peak of \objectname[]{B2250+194} at other
frequencies \citep[see also,][]{Kovalev+2008}. The location of the
peak at the highest radio frequencies is expected to be close to the
source's core and the putative black hole there. For the similarly
compact and elongated source, M81$^*$, in the center of the nearby
galaxy M81, this method did lead to the confirmation of the
approximate location of the core of M81$^*$.  This location was
earlier determined as the most stationary in the varying brightness
distribution of M81$^*$ relative to another source (the shell center
of SN 1993J) in the same galaxy (\citealt{BietenholzBR2001}; see also
\citealt{BietenholzBR2004}). However, our observations at 5 and 15 GHz
were not planned for high-precision astrometry and could not be used
for this purpose. Our 5 and 15 GHz images appear to be, respectively,
just larger and smaller versions of our 8.4 GHz images with the
brightness peak remaining in the center of the image (Paper II),
without giving any hint as to whether the core may be located in the
south of the structure.

As reported in Paper II, \objectname[]{B2250+194} undergoes slight
structure changes. When the source is modeled with an elliptical
Gaussian, the major axis of the Gaussian varies between $\sim$0.6 and
$\sim$0.8 mas along a position angle of $-11$\arcdeg\ over the time from
1997 to 2005. These structural changes could at least partly account
for the nominal proper motion of C1 relative to
\objectname[]{B2250+194}.

What is the $1\sigma$ upper limit on the proper motion of C1?  From
the point of view of our IM Peg VLBI observations from 1997 to 2005
for which we use C1 as a reference, the degree of stationarity for
that period is relevant. For our longest period for the data corrected
with JPL, from 1998 to 2005, we obtain for C1 a 1$\sigma$ proper
motion limit relative to B2250+194 of 38 and 35 \muasyr\ in \al\ and
\de, respectively (solution \#1, Table~\ref{t5comp}).
The equivalent upper limit of the proper motion in the CRF over that
period is 46 and 56~\muasyr\ in \al\ and \de, respectively (solution
\#4, Table~\ref{t2icrf}).  For \GPB\ (see \S~\ref{sintro}), this is 
our fundamental result on
the level of stationarity of the chosen single reference point, C1,
relative to the distant universe.

For the shorter period, from 2002 to 2005, no significant proper
motion for C1 was found above the 1.5$\sigma$\ level, neither
relative to B2250 +194, nor to B2252+172, nor within the CRF.  For
this period we obtain smaller limits that are interesting from an
astrophysical point of view. C1 is stationary relative to the
combination of the two reference sources within a $1\sigma$ upper
limit of 33 and 19 \muasyr\ (solution \#3 in Table~\ref{t2icrf}) and
in the CRF within an upper limit of 39 and 30~\muasyr, in \al\ and
\de, respectively (solution \#5, Table~\ref{t2icrf}).

\subsubsection{Motion of the Jet Components of 3C 454.3}
\label{jetmotion}

What are the proper motions of the components other than C1 relative
to the two reference sources and in the CRF?  Are they significantly
different from that of C1 and, moreover, can significant motion be
detected of any of them relative to the distant universe?  In Paper II
we showed that C2, D1, D2, J1, and Jext are all moving on average away
from C1.
Their relative speeds are not necessarily constant over the more than 7 years
of our observations. Most dramatically, J1 moves at about twice its
average speed until 2002 and then slows down to almost
zero velocity thereafter (Paper II).

In Table~\ref{t5comp} we list the proper motions of these components
relative to each of the two reference sources and to both combined,
and in Table~\ref{t2icrf} to the CRF.
For C2 we find motion with a larger significance than that for C1 only
relative to B2252+172 (solution \#2) and to both reference sources
combined (2.0$\sigma$ in \de, solution \#3). However, we do not
regard the significance large enough to consider the motion real.
Relative to B2250+194 and to the CRF, C2 is stationary at about the
same significance level as C1. For J1 a clear average northward motion
of 49$\pm9$ \muasyr\ (5.4$\sigma$) relative to B2250+194 is observed
between 1998 and 2005 (solution \#1). The significance, however,
decreases to only $1.9\sigma$, due to larger uncertainties, when the
motion is measured in the CRF (solution \#4), comparable to the 
significance of the motions in the CRF of each, C1 and C2.  And the motion itself
decreases to (almost) zero  within the errors for the period from 2002 to
2005 relative to either B2250+194 (solution not listed) or B2252+172
(solution \#2) or both (solution \#3) or to the CRF (solution \#5)
consistent with J1's motion
relative to C1.  The largest and/or most significant motions, relative
to either of the two reference sources or both combined or in the CRF
were found for D1, D2, and Jext, with, for instance, D2 having a speed
in the CRF of $-99\pm25$ and $185\pm59$ \muasyr\ in \al\ and \de,
respectively, for the short period from 2004 to 2005 (solution \#5).

To summarize: For the long period from 1998 to 2005, C1 and C2 are
stationary within the same small bounds relative to B2250+194 and are
joined by J1 when the stationarity is measured (less accurately) in the CRF. The other
components are stationary only within larger bounds or move
significantly.  For the short period from 2002 to 2005, C1 and J1 are
stationary within the same small bounds relative to B2252+172 and to
both reference sources combined, and are joined by C2 when the
stationarity is measured in the CRF. Again, the other components are
stationary only within larger bounds or show significant proper
motion.  The largest significant proper motion was found for D2.

\section{Astrometric Results (3): Analysis of Other Motions}
\label{ast3}

\subsection{Fit: Parallax, proper acceleration, and orbital motion of the core component, C1, of 3C 454.3}

Since our computation of the motion of the guide star
\objectname[]{IM Peg} includes solutions for parallax and proper
acceleration, to help us put limits on certain systematic errors we
also solve for these parameters for C1. In Table~\ref{t6res2} we list
the proper motion and the parallax, $\pi$, obtained for C1 relative to
\objectname[]{B2250+194} and separately to \objectname[]{B2252+172} by
assuming that both of the reference sources are infinitely distant
from Earth.
In addition, we solve for the acceleration components, \dotmua\ and
\dotmud, for C1 and also list these results in Table~\ref{t6res2}.

\begin{deluxetable}{l c c c c c}
\tabletypesize{\small}
\tablecaption{Proper motion, parallax, and proper acceleration\tablenotemark{a}}
\tablewidth{0pt}
\tablehead{
\colhead{Source-reference} 
                 & \colhead{\mua}  & \colhead{\mud}  & \colhead{$\pi$}
                 & \colhead{\dotmua}  & \colhead{\dotmud}              \\
                 & \colhead{(\muasyr)}  & \colhead{(\muasyr)}   & \colhead{(\muas)}
                 & \colhead{(\muasyryr)}  & \colhead{(\muasyryr)}                  
}
\startdata
C1$-$2250          &  29$\pm$\phn8 &$-27\pm\phn8$        &$\phn40\pm20$  &  \nodata  &  \nodata  \\ 
C1$-$2252          &  21$\pm$18    &\phn\phn$4\pm14$     &    $-2\pm17$  &  \nodata  &  \nodata  \\
\\
C1$-$2250          &  33$\pm$26    &\phn\phn$11\pm27$    &$\phn38\pm20$  & \phn\phn2 $\pm$ 9   & 13$\pm$ 9    \\
C1$-$2252          & \phn3$\pm$44  &\phn\phn$13\pm37$    &   $-2\pm18$   & $-21\pm47$    & 11$\pm$ 40   \\

\enddata 
\tablenotetext{a}{Parameter estimates for component C1
of 3C~454.3 relative to \objectname[]{B2250+194} and, separately,
to \objectname[]{B2252+172}. 
Uncertainties are statistical standard errors derived from the weighted
least-squares fit, scaled to $\chinu = 1$. The reference epoch is 2005
February 1.} 

\label{t6res2}
\end{deluxetable}

Our two most accurate solutions for parallax (relative to B2252+172),
unsurprisingly, are zero (to within $0.1\sigma$), with the upper limit being
$\sim$20~\muas, corresponding to a distance, $D$, of $>50$~kpc. Although 
other data sets might well yield more accurate parallax measurements, 
or bounds thereon, ours is one of the most, if not the most, accurate so far obtained, 
given the assumption that the reference sources are sufficiently 
distant to have negligible parallax. The
proper acceleration of C1 relative to B2250+194 is within 1.4$\sigma$
of zero, and is not significant.

We also extended the number of free parameters still further and included a
fit to orbital parameters corresponding to the period of the IM~Peg
binary system of 24.6~d, since such a fit is used in our analysis
of IM Peg (Paper V)\@.  We found no indication of such orbital motion
of C1, with each of the orbital parameters being zero within 1$\sigma$.

\subsection{Non-linear motion of the core component, C1, of 3C 454.3 on the sky?}

\label{nonlinearmotion}

\begin{figure}
\centering
\includegraphics[width=\textwidth]{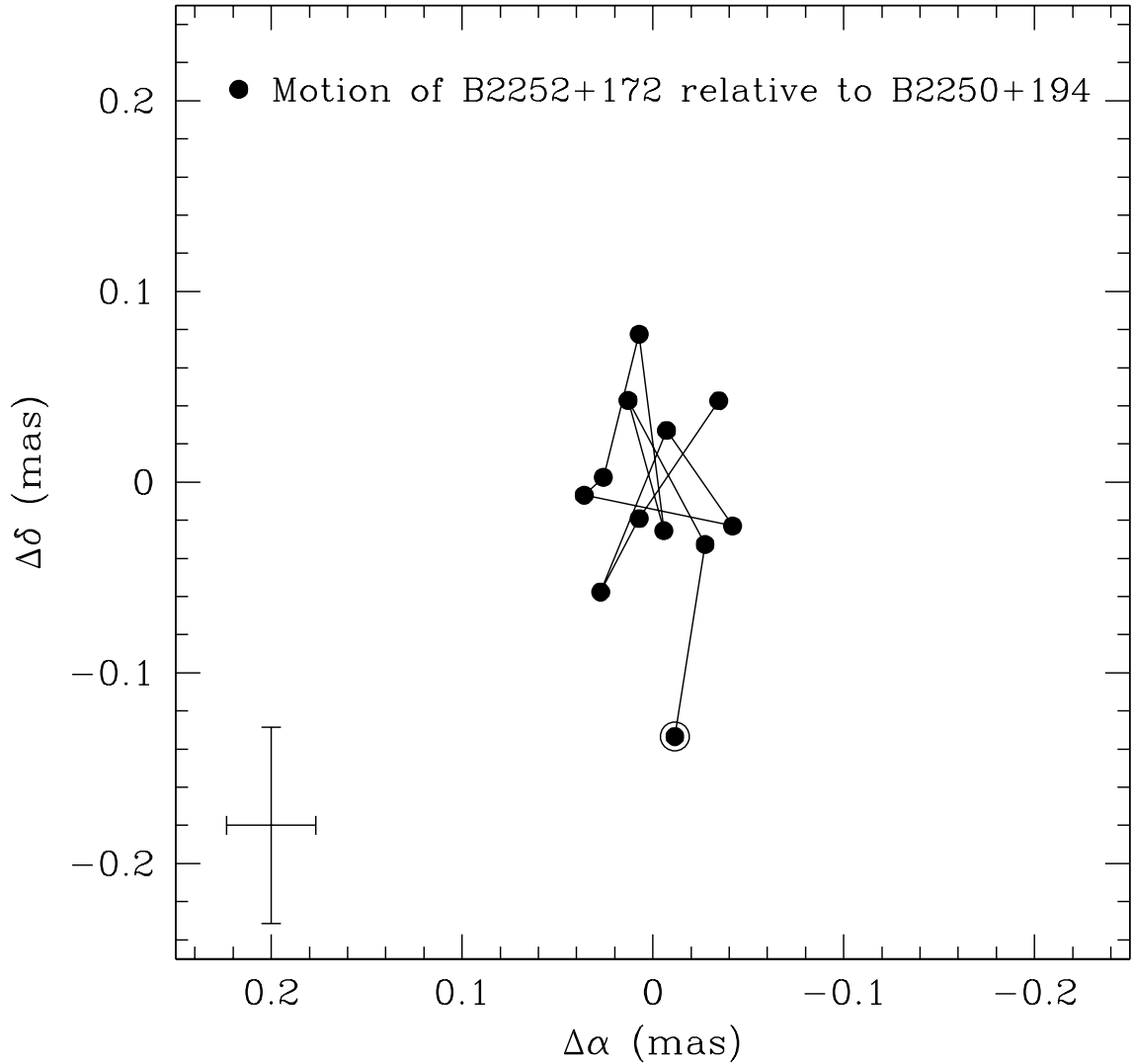}
\caption{The coordinate determinations, except for an offset, of
B2252+172 relative to those of B2250+194 obtained by differencing the
values in Tables~\ref{t32250} and \ref{t42252} for each epoch in the
sense (C1$-$2250)$-$(C1$-$2252).  Errors are left unplotted
for clarity but can be computed from the errors given in the above
tables. The data point for the first epoch, 2002 Nov.\ 20, is
indicated by a circle around the black dot. The cross in the lower
left indicates the weighted rms of the scatter of the data points in the figure, of
23~\muas\ in \al\ and 51~\muas\ in \de.}
\label{f1550x52ijpl}
\end{figure}

\begin{figure}
\centering
\includegraphics[width=\textwidth]{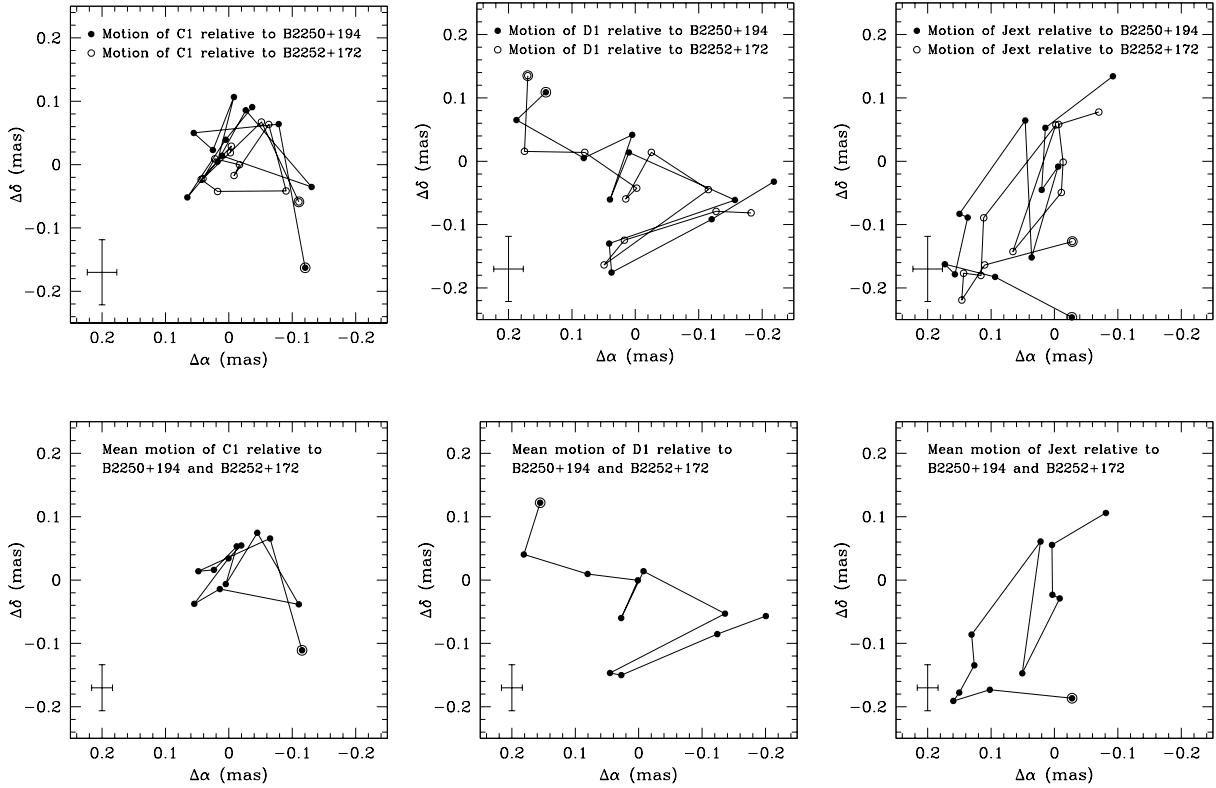}
\caption{The coordinate determinations of C1, D1, and Jext, except for
an offset for each, relative to both B2250+194 and B2252+172 for each epoch
(upper row). The values were taken from Tables~\ref{t32250} and
\ref{t42252}.  The lower row gives the coordinate determinations as
averages in the sense, (C1$-$2250)/2 + (C1$-$2252)/2.  The data
points for the first epoch, 2002 Nov.\ 20, are indicated in the upper panels by a circle
around (i) the black dot for the motions relative to B2250+194  
and (ii) the open circle for the motions 
relative to B2252+172; in each of the lower panels the corresponding circle is again around a black dot. The crosses in
the lower left of each upper panel are taken from
Fig.~\ref{f1550x52ijpl} and approximately indicate the standard errors
of the data points.  For the lower panels, the standard errors are
assumed to be smaller by $\sqrt[]{2}$.}
\label{f16c1d1j2x5052}
\end{figure}

Is there significant motion of the core component, C1, of 3C~454.3 on
the sky that departs from the (linear) proper motion inherent in our
fit models?  Inspecting Figure~\ref{f9c15052}, we see that the motion of
C1 relative to B2250+194 is somewhat correlated with C1's motion
relative to B2252+172 in both \al\ and \de.
To investigate possible non-linear, or in general any unmodeled,
motion in more detail we first plot the coordinates of B2252+172
relative to those of B2250+194 in Figure~\ref{f1550x52ijpl}. We are
using only data corrected for ionospheric effects with the JPL model
since this model seems superior to the PIM model and since we are only
using data taken during the period of our B2252+172 observations for
which corrections with the JPL model were available, i.e. from 2002 to
2005. The plot shows quasi-random motion with wrms values of 23~\muas\
in \al\ and 51~\muas\ in \de.  In Figure~\ref{f16c1d1j2x5052} (upper
left panel), we plot the positions of C1 relative to those of B2250+194
and B2252+172.  It is apparent that the positions of C1 are covering a
larger area than the positions of B2252+172 in the previous figure,
mainly because of a larger scatter along \al. The peak-to-peak
variations are 2.5 times larger in \al\ but only 1.2 times larger in
\de\ than in the preceding figures.
Also, the positions of C1 relative to those of B2250+194 resemble the
positions of C1 relative to those of B2252+172. These two indications
may be evidence for C1 apparently moving on the sky along the east-west
axis above the noise level which we adopt to be the wrms values from
the B2252+172 versus B2250+194 plot and indicate by the cross in the
lower left corner of the figure.  In the lower left panel we plot the
mean of each pair of position determinations (with $\sqrt[]{2}$
smaller cross bars) to display this apparent motion more clearly. The
plotted motion is confined to an area not larger than $\sim$0.2
$\times$0.2 mas$^{2}$, a small portion of the beam area but larger
than the area of ``jittery'' motion in Figure~\ref{f1550x52ijpl}.  Any
east-west jittery motion of C1 is consistent with our simulations (Paper
II) of how C1 moves east-west relative to the larger
core region structure and relative to the 43-GHz core located $0.18\pm0.06$~mas
east of C1. This east-west motion simulated for C1 is confined to within
0.12 mas. Our finding of possible motion within 0.2 mas is less
precise than the result from Paper II, but in principle more accurate
since it is measured relative to physically unrelated sources nearby
on the sky.

For comparison we show the equivalent motions of the components D1 and
Jext in the middle and right columns of Figure~\ref{f16c1d1j2x5052}.  As
expected from our solution for the proper motion, D1's motion has a
linear component to the west-southwest and Jext's toward the
northwest.

\section{Discussion}
\label{discuss}

\subsection{Considerations for geodetic VLBI with group delays}

By comparing the positions of C1 and 3C~454.3, both in the CRF, we
found a possibly significant difference of $-338 \pm 105$~\muas\
(3.2$\sigma$) in \de\ while the difference in \al\ is not
significant. To repeat, the position of C1 in the CRF was determined from the
position of B2250+194 in the CRF by the addition of the position of
C1 relative to that of B2250+194. Both the positions of 3C 454.3 and
B2250+194 in the CRF were determined from geodetic VLBI group delays
while the relative position (C1$-$2250) was determined from our
differential VLBI phase delays. In contrast to the latter, the
group-delay observations were not corrected for structure
effects. While such effects can be assumed to be insignificant for
B2250+194, they may well be significant for 3C~454.3. Therefore, in
contrast to that for B2250+194, the position solution for 3C~454.3
does not refer to a particular reference point in the source's
brightness distribution. The solution should therefore be influenced
by the structure of the source and its changes with time and
frequency.  In particular, a shift of the position of the core with decreasing frequency 
away from the foot of the jet but along the jet axis \citep[see, e.g.,][]{Sokolovsky+2011} could influence the results from group 
delays and phase delays differently.

But the source is largely oriented east-west in its
brightest part, namely the core region, and oriented toward the
northwest in its low-luminosity 10-mas long jet. On first sight one might
therefore expect a larger discrepancy in \al\ \citep[see,
also,][]{Porcas2009} and a smaller one, if any, in \de\ and then with
a shift to the north, not as observed to the south. If the 10-mas jet caused 
the discrepancy, it would be via 
some peculiar influence.  The group-delay
position determined in 2007 at 24 GHz \citep{Lanyi+2010} 
indeed diminishes the \de\ discrepancy in the
position of C1 by about half although the uncertainty of the 24-GHz position
is large. In these observations, 3C~454.3 appears more compact than it does at 8.4
GHz.  The most easterly component at 24 GHz appears to be point-like
and dominates the image. Low-brightness extended features appear up to
3 mas toward the west but the 10-mas jet is not visible
\citep{Charlot+2010} and should therefore have essentially no
influence on the position determination.  In this context,
\citet{Marti-Vidal+2008} have also found apparently significant
discrepancies between source positions determined from geodetic VLBI
group delays without structure corrections and those determined from
differential VLBI phase-delays with structure corrections. All said,
the cause of the discrepancy is not understood.

\subsection{Astrophysical implications}
\subsubsection{Limit on the proper motion of the core of 3C 454.3 and the
proper motions of the jet components}

Since the component C1 is as close to stationary in the CRF as is any
of our other five components of 3C~454.3., and could be closely
related to the easternmost compact flat-spectrum component in 43-GHz
images (Paper II), this component is likely the closest in our images
to the expected supermassive black hole, and therefore very near the
center of mass of the quasar.  The radio emission probably originates
close to the foot of the jet in the vicinity of this putative black
hole.  The jet components, C2, D1, D2, J1, and Jext are all moving
away from C1 (Paper II). The motions of D1, D2, and Jext are also
significant in the CRF at our sensitivity levels.  The proper-motion
values of all six components of 3C~454.3 from Table~\ref{t5comp}
(solution \#5) were converted to apparent velocities and are listed
in Table~\ref{t7vel}.  Our $1\sigma$ upper limit on the proper motion of C1 in the CRF
of 39 and 30~\muasyr\ in \al\ and \de, respectively, corresponds to a
limit of 1.0 and 0.8 $c$.  The speeds of the jet components in the CRF
can be faster and superluminal; for instance for D2 the speed is
$\sim5 c$.

\begin{deluxetable}{c@{\hspace{0.3in}} c@{\hspace{0.3in}} c@{\hspace{0.6in}} c}
\tabletypesize{\small}
\tablecaption{Velocity of the components of 3C 454.3 in the CRF\tablenotemark{a}}
\tablewidth{0pt}
\tablehead{
\colhead{Component}      &      \colhead{$v_{\alpha}/c$}      &
\colhead{$v_{\delta}/c$} & \colhead{Range of epochs\tablenotemark{b}}
}
\startdata
C1                  &   $\phn0.6\pm0.4$     & $0.1\pm0.7$ & 2002.89 -- 2005.54  \\
C2                  &   $   -0.2\pm0.4$     & $0.6\pm0.7$ & 2002.89 -- 2005.54    \\
D1                  &   $   -3.2\pm0.6$     & $2.3\pm0.7$& 2003.07 -- 2005.54 \\
D2                  &   $\phn2.5\pm0.6$     & $4.7\pm1.5$& 2004.18 -- 2005.54  \\
J1                  &   $\phn0.3\pm0.4$     & $0.3\pm0.6$& 2002.89 -- 2005.54  \\
Jext                &   $   -1.5\pm0.6$     & $2.5\pm0.7$& 2002.89 -- 2005.54  \\
\enddata
\tablenotetext{a}{Velocity in $\alpha$ and $\delta$ in units of the
speed of light for angular velocities in Table~\ref{t2icrf} (solution
\#5). The velocities were computed for an angular diameter distance of
3C~454.3 of 1.6 Gpc.}
\tablenotetext{b}{Same as in Table~\ref{t5comp}.} 
\label{t7vel}
\end{deluxetable}

Similar characteristics are displayed by the superluminal quasar,
3C~345.  The core was found to be stationary relative to the
physically unrelated quasar, NRAO 512, within $0.4 c$ in the east-west
direction while the jet components moved away from the core in this
direction at up to $9 c$ \citep{Bartel+1986}\footnote{We assume the
same cosmological parameters as we use here for 3C~454.3.}.
Since in each of these two cases the core is compact and has an
approximately flat spectrum in the GHz frequency range, we conclude
that compactness of the component and flatness of the spectrum are
indeed, as generally assumed, indicative of the nearby presence of the
gravitational center of the quasar.

Component J1 is special partly because the limit on
its speed over the short time interval from 2002 to 2005 is subluminal and partly
because its speed was likely much larger at earlier times.
\citet{Pauliny-Toth1998} found from VLBI observations at 11 GHz that a
component, dubbed ``A,'' moved away from the core toward the northwest
from a distance of $\sim$2.8 mas in 1984 to a distance of $\sim$5.0
mas in 1992 with a speed averaging $25 c$, but varying from $15 c$ between 
1984 and 1985 to
$30 c$ between 1985 and 1989 and $20 c$ between 1989 and 1992.
This component is likely to be our J1.  \citet{Jorstad+2005} found
that by 2001 this component, dubbed ``D'' in their paper, has possibly
decelerated considerably to $6 c$.
The strong deceleration can also be seen in our data from 1998 onward
such that by 2002 to 2005 the component remained stationary within our
subluminal limits (see Table~\ref{t7vel}).

Larger speeds of up to $530\pm50$~\muasyr\ were reported for jet
components that could be discerned with higher angular resolution at
43 GHz in the C1 and C2 area \citep{Jorstad+2001b}.  These authors
also speculated on whether perhaps the 43-GHz core could have moved to
the northeast by 0.1 mas in \al\ and 0.2 mas in \de\ between 1997.6
and 1998.2. While this time range is just before the start of our VLBI
observations, our proper-motion measurements limit any such motion for
later times.

Recently an exceptionally bright optical outburst was detected in
3C~454.3 \citep{Villata+2006, Vercellone+2010} reaching a maximum in spring
2005. It was accompanied by an increase of radio emission at 43 GHz
from the core that started in early 2005 and reached a maximum in
September 2005. If the outburst is associated with activity of the
core, perhaps with the ejection of a new jet component, then the
position of C1 may be expected to be affected.
However, inspection of our graphs of the temporal changes 
in the position of C1 does not show any indication of a possible
emerging jet component near C1, a result not surprising 
given the 2005 mid-July end of our data set.

To repeat, any transverse linear motion on the sky found for the core
of 3C~454.3 has a speed $\leq 1.0 c$. This limit is almost as low in magnitude
as the radial velocity of the quasar, computed from its
redshift. Transverse velocities comparable in magnitude to the
redshift velocities are not expected for quasars for a cosmological
model where the dominant motion is the redshift velocity due to the
expansion of the universe.
Any transverse velocity for the cores of quasars should be at least
two orders of magnitude smaller than $c$ based on our knowledge of
peculiar motions of galaxies and galaxy clusters.
The effect of acceleration of the solar-system barycenter toward the
galactic center is also expected to be relatively small.
However, it is plausible that a non-zero mean proper motion with
respect to the CRF of physically unrelated sources that are separated
by less than, say, a radian on the sky is significant because, in
effect, we construct the CRF using the approximation that the position
of the currently estimated solar-system barycenter is an inertial
reference.  Fortunately, the estimated acceleration of that barycenter
toward the Galactic center has a relatively small effect: At its
maximum value (approximately applicable for our sources), the effect
is only about 4~\muasyr\ \citep{Sovers+1998, Titov2010} and has recently been reported
to be observed \citep{Titov+2011}.

Moreover, any unexpectedly
large acceleration of the solar-system barycenter is less likely, given 
the study by \citet{ZakamskaT2005}, who find that pulsar timing
data (from both single and binary pulsars) are inconsistent with any
unmodeled accelerations of the solar-system barycenter greater than
$\sim4\times 10^{-9}$ cm s$^{-2}$, which is only about twice the magnitude of
the galactocentric acceleration.  The more recent pulsar VLBI results
of \citet{Deller+2008} for PSR J0437-4715 likely strengthen this upper limit.

In the future it may be possible to search for the proper motion of
the cores of quasars with uncertainties much smaller than $c$. Then it
could be confirmed for the first time that the Hubble flow 
dominates the motion of quasar cores and that the velocity expected
from the solar-system-barycenter acceleration is indeed to a high
degree consistent with models.

\subsubsection{Non-linear motion of the core within the boundaries of the proper-motion limit}

Our observation of possibly significant jittery motion of the core
within an area of the sky as small as 0.2$\times$0.2 mas$^{2}$ would
be only the second time such motion has been recorded unambiguously
for a source with core-jet structure by using as a reference for the
motion a physically unrelated source nearby on the
sky. The first source for which such motion was detected is the core
in the core-jet structure of the nearby galaxy M81. In this case, the
center of the expanding shell of the nearby supernova 1993J was used
as a reference \citep{BietenholzBR2001}.

What caused this apparent non-linear motion?  Could such motion be
indicative of orbital motion related to a binary black-hole system?
This possibility is unlikely since the motion is jittery and its
magnitude too large to be physically reasonable.

Our measurements of the jittery motion of C1 along the north-south
direction is within the noise determined from the jitter in the
positions of B2250+194 relative to B2252+172.  C1's jittery motion
along the east-west direction is above the noise level.  In fact, the
peak-to-peak variation is 2.5 times larger than the corresponding
variation in the positions of the two reference sources and may be
significant.  If so, it is likely caused by slight
brightness-distribution changes due to activity at the foot of the jet
close to the putative supermassive black hole.  Such changes may have
influenced the fit position of the Gaussian core component, C1 (Paper
II).

This result, if confirmed, has implications for high-precision astrometric
observations in general.  It shows that any component, even a core
component, clearly identified in the structure of a celestial object
may still move on the sky.

\subsection{The relevance for \GPB}

The goal of \GPB\ at launch was to measure the precession of the
gyroscopes relative to distant inertial space with a standard error of
0.5 \masyr\ or less in each sky coordinate. To be a minor
contributor to the error budget, the proper motion of the guide star,
\objectname[]{IM Peg}, was to be determined with a standard
error no larger than 0.14~\masyr\ in each sky coordinate.  Our reference 
source, the quasar
3C~454.3 was shown to be stationary within the CRF over $\sim30$ years of
geodetic VLBI observations to within 0.023 and 0.009~\masyr\ in \al\
and \de, respectively. More to the point, our primary reference point
for \GPB, C1 in 3C~454.3, is stationary with respect to B2250+194 to
within 0.038 and 0.035 \masyr\ and within the CRF to within 0.046 and
0.056~\masyr\ in \al\ and \de, respectively, over the seven years that almost
entirely cover the period of our VLBI observations in support of the
\GPB\ mission and is therefore a negligible contributor to the error
budget of the proper motion of the guide star.

\section{Conclusions}
\label{conclus}

\noindent Here we summarize our observations and data analysis, and give our conclusions:

\begin{trivlist}
\item{1.} We made differential VLBI observations at 35 epochs of the
quasar \objectname[]{3C 454.3} and the radio galaxy
\objectname[]{B2250+194} along with 12 epochs of the extragalactic,
unidentified source, \objectname[]{B2252+172}, at 8.4 GHz between 1997
and 2005.  With these sources we provided a reference frame composed
of extragalactic sources nearby on the sky to \objectname[]{IM Peg}
and, together with geodetic VLBI observations made by others we
provided a (global) CRF, the latter closely linked to the ICRF2, 
for the determination of the proper motion of
the \GPB\ guide star IM Peg with respect to the distant universe.

\item{2.} We analyzed our differential VLBI observations using
phase-referenced mapping and phase-delay fitting in combination with a
Kalman filter.

\item{3.} Our $1\sigma$ upper limit of the proper motion of \objectname[]{B2252+172} relative to
\objectname[]{B2250+194} is $11$~\muasyr\ in \al\ and
$24$~\muasyr\ in \de\ for the time from 2002 to 2005, identifying
\objectname[]{B2252+172} unequivocally as extragalactic and providing
for a highly stable reference frame of sources nearby on the sky to
\objectname[]{3C 454.3}.

\item{4.} Our $1\sigma$ upper limits of the proper motions of \objectname[]{3C 454.3}, 
\objectname[]{B2250+194}, and \objectname[]{B2252+172} in the CRF determined with geodetic observations,
and the latter two also partly with phase-delay observations, are $\leq30$~\muasyr\ in each coordinate.

\item{5.} Our $1\sigma$ upper limit on the proper motion of C1, our
core component of 3C~454.3, relative to 
the combination of B2250+194
and B2252+172 for the time from 2002 to 2005 is $<35$ \muasyr\ in each of \al\
and \de, indicating that C1 is
highly stable with respect to two extragalactic sources
nearby on the sky.

\item{6.} The 1$\sigma$ upper limit on the proper motion of C1 in the CRF for the
time from 1998 to 2005 is 46 and 56 \muasyr\ in \al\ and \de, respectively.. This is our
fundamental result of the stationarity of the reference point for the
guide star IM Peg in support of the \GPB\ mission.  For the shorter
time from 2002 to 2005 the 1$\sigma$ upper limit on the proper motion
of C1 in the CRF is 39 and 30 \muasyr\ for the two coordinates,
respectively, corresponding to subluminal motion of $\leq
1.0 c$ and $< 0.8 c$, for \al\ and \de, respectively, for an
angular-size distance to 3C~454.3 of 1.6 Gpc, for a flat universe with
$H_0$=70~\kmsMpc, $\Omega_M=0.27$, and $\Omega_{\lambda}=0.73$.

\item{7.} The source  coordinates of C1 in the CRF 
differ from those of 3C~454.3 determined from geodetic group-delay data
by $131\pm98$~\muas\ in \al\ and $-338\pm105$~\muas\ in \de, the latter coordinate being different at 
the 3.2$\sigma$  level. This difference in \de\ is not understood, except possibly as a statistical fluke.

\item{8.} C2 and J1, the latter for the period from 2002 to 2005, are 
stationary in the CRF to within bounds as small as those for C1.
However, C1's
flat spectrum and compactness in contrast to the spectra and the compactness of
the other components, indicate that C1 is closest
to the putative supermassive black hole and the probable gravitational
center of the quasar.

\item{9.} The jet components, D1, D2, and Jext clearly
move in the CRF.
Their motions correspond to superluminal speeds, which for D2 is $5~c$.

\item{10.} Notwithstanding our limit on the proper motion of C1, there
is evidence for its having jittery $\sim0.2$~mas east-west motion
above the noise level, likely related to jet activity in the vicinity
of the core. This evidence is consistent with the jittery motion of C1
found in Paper II.
 
\item{11.} The 1$\sigma$ upper limit on the parallax of C1 relative to
\objectname[]{B2250+194} and B2252+172 is 20~\muas, one of the most, if not the most, accurate
limit so far obtained, corresponding to an unsurprising lower limit of 50 kpc
on its distance from Earth and demonstrating the sensitivity of parallax measurements with VLBI. 

\item{12.}  The upper limit on the proper motion of 3C~454.3 over $\sim30$
years of geodetic VLBI observations and of C1 over $\sim8.5$ years of
our phase-delay VLBI observations
is sufficiently small to meet the goal of the \GPB\ mission and
therefore to justify use of C1 as the primary reference point for
\GPB.
\end{trivlist}

ACKNOWLEDGMENTS.  This research was primarily supported by NASA,
through a contract with Stanford University to SAO, and a subcontract
from SAO to York University.  The National Radio Astronomy Observatory
(NRao) is a facility of the National Science Foundation operated under
cooperative agreement by Associated Universities, Inc.  The DSN is
operated by JPL/Caltech, under contract with NASA\@.  This research
has made use of the United States Naval Observatory (USNO) Radio
Reference Frame Image Database (RRFID)\@.  We have made use of NASA's
Astrophysics Data System Abstract Service, developed and maintained at
SAO\@. Jeff Cadieux and Julie Tome helped with the data reduction
during their tenure as students at York University.  We thank the
International VLBI Service for Geodesy and Astrometry
\citep[IVS;][]{SchluterB2007} for their support.


\end{document}